\documentclass{emulateapj}
\usepackage{graphicx}
\usepackage{epigraph}
\slugcomment {To appear in the Astrophysical Journal }
\shorttitle{Spectra of SNR 1885 in M31}

\shortauthors{Fesen et al.}

\newcommand{\dd}{d}
\newcommand{\unit}[1]{\, {\rm #1}}
\newcommand{\Msun}{\unit{M}_\odot}

\newcommand{\masstable}{
    \begin{deluxetable*}{lll}
    \tablewidth{0pt}
    \tablecaption{
    \label{masstable}
    Model Parameters
    }
    \tablehead{\colhead{Parameter} & \colhead{This paper} & \colhead{\cite{Fesen1999}}}
    \startdata
    \ion{Fe}{1} mass & $0.05 \Msun$ & $0.013 \Msun$ \nl
    \ion{Fe}{2} mass & $\sim 0.5 \Msun$ & $\sim 0.2 \Msun$ \nl
    \ion{Ca}{1} mass & $0.00024 \Msun$ & $0.00029 \Msun$ \nl
    \ion{Ca}{2} mass & $0.0038 \Msun$ & $0.005 \Msun$ \nl
    \ion{Mg}{1} + \ion{Mg}{2} mass & $0.03 \Msun$ & \nodata \nl
    Maximum expansion velocity & $13{,}400 \unit{km} \unit{s}^{-1}$ & $13{,}100 \unit{km} \unit{s}^{-1}$ \nl
    Covering fraction of Fe & 0.38 & 1 \nl
    Minimum Ca expansion velocity & $2{,}000 \unit{km} \unit{s}^{-1}$ & $0$ \nl
    Minimum Mg expansion velocity & $7{,}000 \unit{km} \unit{s}^{-1}$ & \nodata \nl
    Foreground starlight fraction & $0.16$ & $0.21$
    \enddata
    \end{deluxetable*}
}

\newcommand{\snmasstable}{
    \begin{deluxetable}{lll}
    \tablewidth{0pt}
    \tablecaption{
    \label{snmasstable}
    Masses of Elements in SN Models
    }
    \tablehead{
\colhead{} & \colhead{5p02822.16} & \colhead{5p02822.25} \nl
\colhead{Element} & \colhead{($\Msun$)} & \colhead{($\Msun$)}
    }
    \startdata
He              & $9.54 \times 10^{-4}$ & $0.00116$ \nl
C               & $0.0156$              & $5.44 \times 10^{-4}$ \nl
O               & $0.117$               & $0.0503$ \nl
Mg              & $0.0322$              & $0.0102$ \nl
Si              & $0.399$               & $0.206$ \nl
S               & $0.218$               & $0.149$ \nl
$^{20}$Ne       & $0.00628$             & $0.00167$ \nl
$^{24}$Mg       & $0.0322$              & $0.0103$ \nl
$^{36}$Ar       & $0.0410$              & $0.0356$ \nl
$^{40}$Ca       & $0.0387$              & $0.0400$ \nl
$^{56}$Ni       & $0.315$               & $0.601$ \nl
Fe              & $0.409$               & $0.699$
    \enddata
    \end{deluxetable}
}

\begin{document}

\title{Optical and UV Spectra of the Remnant of SN~1885 (S And) in M31\altaffilmark{1}}

\author{Robert A.\ Fesen\altaffilmark{2}, 
        Kathryn E.\ Weil\altaffilmark{2},
        Andrew J. S. Hamilton\altaffilmark{3}, and
        Peter A. H\"oflich\altaffilmark{4}
        }
\altaffiltext{1}{Based on observations with the NASA/ESA Hubble Space Telescope,
obtained at the Space Telescope Science Institute,
which is operated by the Association of Universities for Research in
Astronomy, Inc.\  under NASA contract No.\ NAS5-26555.}
\altaffiltext{2}{6127 Wilder Lab, Department of Physics \& Astronomy,
                 Dartmouth College, Hanover, NH 03755}
\altaffiltext{3}{JILA and the Department of Astrophysical and Planetary Sciences,
                 University of Colorado, Boulder, CO 80309}
\altaffiltext{4}{Department of Physics, Florida State University, Tallahassee, FL 32306}

\vspace*{\fill}

\begin{abstract}

We present multi-slit, 1D and 2D optical and ultraviolet spectra of the remnant
of Supernova 1885 (SN~1885; S~And) taken using the Hubble Space Telescope's
Imaging Spectrograph (HST/STIS). These spectra of this probable subluminous Type Ia
remnant, seen in silhouette against the central bulge of the Andromeda Galaxy
(M31), show strong and broad absorptions from neutral and singly-ionized
species of calcium, magnesium and iron, but with strikingly different
distributions.  Calcium H \& K absorption indicates spherically distributed
Ca-rich ejecta, densest in a lumpy shell expanding at $2000$ to $6000$ km
s$^{-1}$. Equally broad but weaker Ca~I 4227 \AA \ absorption is seen to extend
out to velocities of $\sim13{,}000$ km s$^{-1}$. Magnesium-rich ejecta in the
remnant is detected for the first time through Mg~I 2852 \AA \ and Mg~II 2796,
2803 \AA \ absorptions concentrated in a shell with expansion velocities from
$\simeq7000 \unit{km} \unit{s}^{-1}$ to at least 10,000 km s$^{-1}$.  Fe~I 3720
\AA \ absorption is detected as two discrete blueshifted and redshifted absorptions
suggestive of an Fe~I shell with expansion velocities of $\pm 2000 - 8000$ km
s$^{-1}$. Weak Fe~II resonance absorptions in the wavelength region 2300 --
2700 \AA \ are consistent with prior {\sl HST} UV images showing \ion{Fe}{2}
rich ejecta confined to a small number of optically thick plumes.  The presence
of such iron plumes extending out from the remnant's core plus layered shells
of calcium and magnesium point to a delayed-detonation explosion.  The spectra
also suggest a roughly spherical explosion, contrary to that expected by a merger or
collision of two white dwarfs. We conclude that SN~1885 likely was an
off-center, delayed-detonation explosion leading to a subluminous SN~Ia similar
to SN~1986G.

\end{abstract}


\section{Introduction}

{\it{The apparition of this star in the center of so conspicuous a nebula
will long afford matter for reflection, and a theme for discussion by savants.
It raises the question whether, after all, there is, even in the heavens,
such a thing as stability. }}

~~~~~~~~~~~~~~~~~~~~~~~~~~~~~~~~~~~~~~~~~~~~~Lewis Swift 1886
\bigskip

Type Ia supernovae (SNe~Ia) are thought to be the explosions of degenerate
carbon-oxygen white dwarfs that undergo a thermonuclear runaway when they
reach the Chandrasekhar limit \citep{hf60,CK69,Nomoto84,HN00,li03}.  Possible
SN~Ia progenitor systems have been discussed by \citet{Howell2011},
\citet{Nugent2011}, \citet{Bloom2012}, \citet{Stefano2012},
\citet{Hoeflich2013}, and \citet{Maeda2016}.

Observed SN~Ia spectra and light curves suggest a layered structure with
intermediate-mass elements on the outside and nickel-iron material on the
inside.  \cite{khokhlov91} proposed that this structure could be explained
empirically by a ``delayed-detonation'' scenario, in which the
explosion starts in the core as a subsonic burning deflagration wave
that at some point transitions into a supersonic detonation.
The initial deflagration preheats and expands the star's outer layers,
so that when the deflagration transitions to detonation,
the burning of the outer layers does not continue to completion
and a layered structure of intermediate-mass elements is produced. 

SN~Ia explosions involve a complex interplay of turbulent hydrodynamics,
nuclear burning, conduction, radiative transfer in iron-group rich material,
and perhaps magnetic fields
\citep{khokhlov95,Neimeyer95,Livne99,Reinecke99,gamezo2004,roepke12}.  This
complexity means that there is substantial uncertainty in modeling SN~Ia
explosions.  Several basic questions about the 
details of SN~Ia explosions could be answered by direct observations of the
distribution and kinematics of supernova ejecta in young SN~Ia remnants.

\begin{figure*}[t]
   \begin{center}
        \includegraphics[scale=.5]{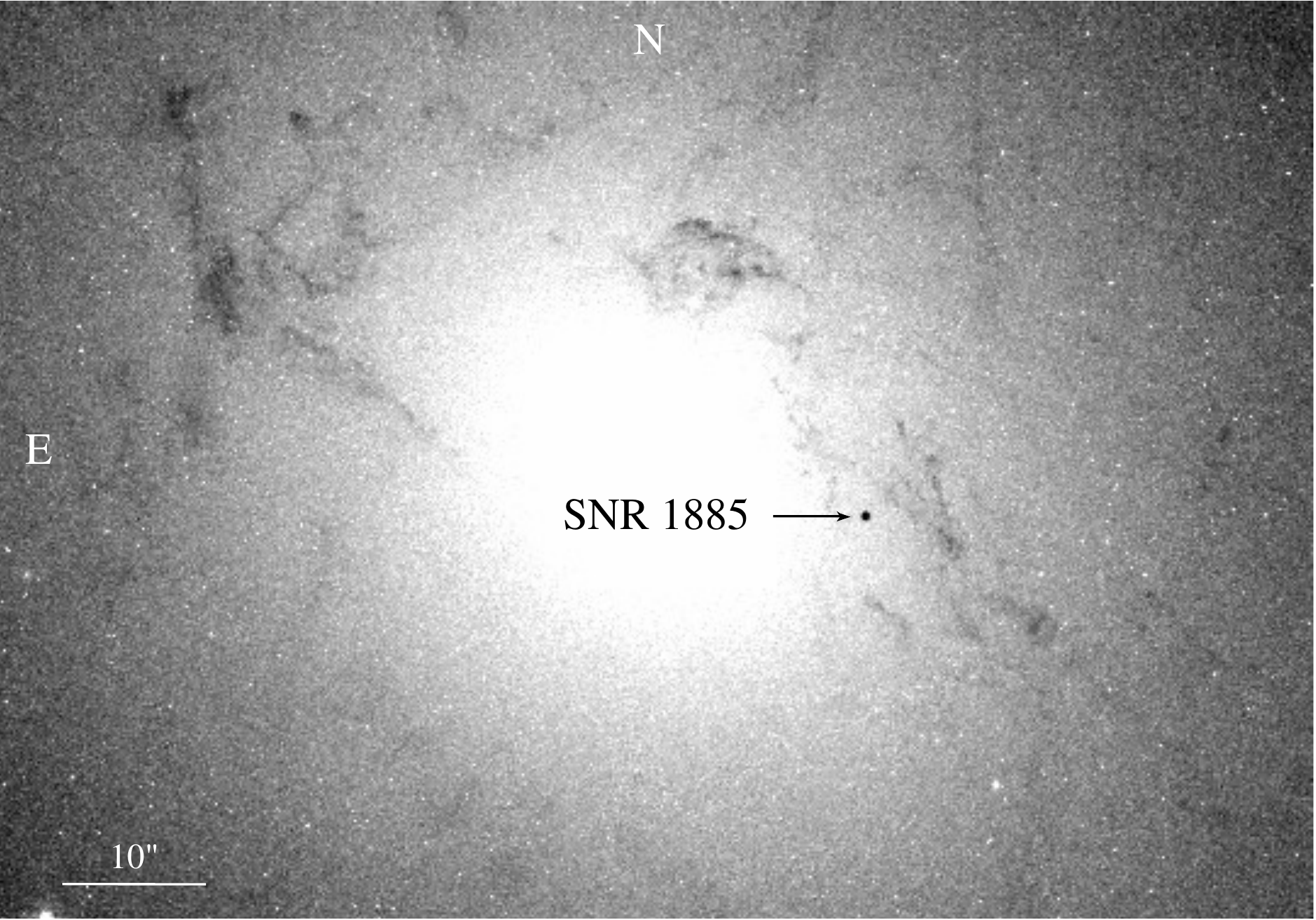}\\
\includegraphics[scale=.5]{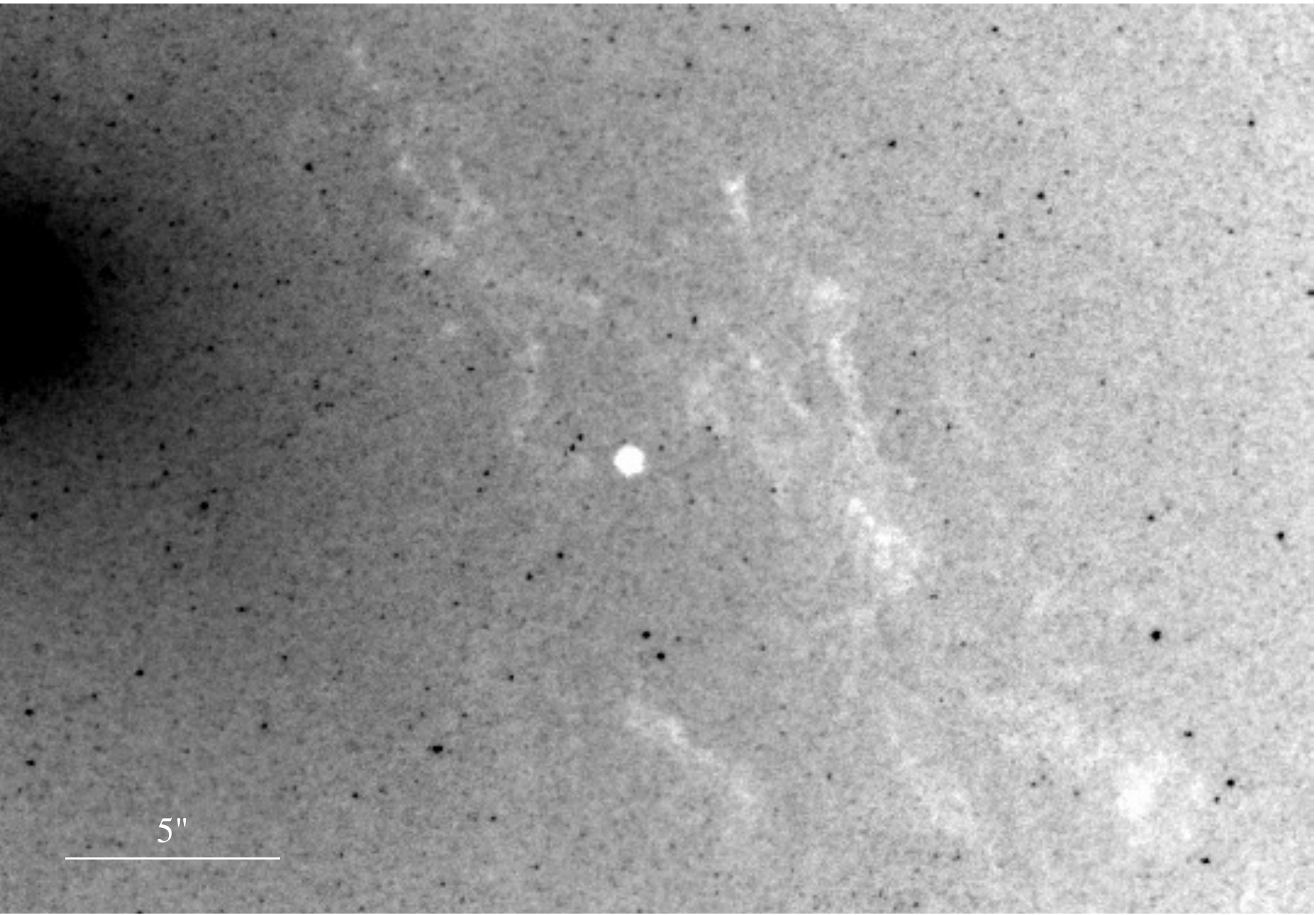}
    \end{center}
        \caption{A December 2010 HST/WFC3 image of the bulge of M31 taken with the F390M filter
as part of an emission-line mapping program of the nuclear regions (PI: Z. Li).
A positive linear stretch of the M31 bulge is shown in the upper panel.
The remnant of SN~1885 appears as a small ($0.8''$) round, dark spot
of \ion{Ca}{2}~H~\&~K absorption $16''$ southwest of the nucleus.
The bottom panel shows an enlarged section of this same image centered on the SN 1885 region
but shown in a negative log stretch (for details, see  \citealt{Fesen2015}). }
        \label{fig:introfigs}
\end{figure*}

The bright historical nova seen in 1885 known as S~Andromeda (S~And; SN~1885),
located $16''$ away from the Andromeda galaxy's (M31) nucleus, offers an
opportunity to study the expanding debris of a SN~Ia. Because of its central
location in M31, the remnant's expanding ejecta are detectable through
resonance-line absorption against the background of M31's bulge stars
(\citealt{Fesen89,Fesen2015}; see Fig.~\ref{fig:introfigs}).

The reported optical spectrum of SN~1885 lacked hydrogen lines, defining it as
Type~I \citep{deV85}.  Its orange appearance, maximum brightness, and fast light
curve are also consistent with a subluminous Type~Ia event (e.g., see
\citealt{Maunder1885,Maunder1886,Huggins1886}). However, this classification
has been questioned \citep{CP88,Pastorello08,Perets2011}.

Images taken with the Hubble Space Telescope ({\sl HST}) in 1995 revealed a
circular $\simeq 0\farcs70 \pm 0\farcs05$  diameter dark spot produced by a blend of
\ion{Ca}{2} H \& K line absorption \citep{Fesen1999}. Subsequent ultraviolet
imaging of the remnant with {\sl HST} detected a low signal-to-noise
$\sim 0\farcs5$ diameter absorption spot attributed to UV \ion{Fe}{2} resonance lines
\citep{Hamilton00}.  

Spectra taken with Faint Object Spectrograph (FOS) on {\sl HST} established
that absorption over $3200$ -- $4800 \unit{\AA}$ was produced principally by
\ion{Ca}{2}~K~\&~H $3934$, $3968 \unit{\AA}$, with additional contributions
from \ion{Ca}{1} $4227 \unit{\AA}$ and \ion{Fe}{1} $3441 \unit{\AA}$ and $3720
\unit{\AA}$ \citep{Fesen1999}.  The remnant's \ion{Ca}{2} absorption was found
to extend to a velocity of $13{,}100 \pm 1500 \unit{km} \unit{s}^{-1}$.  
The strength of the \ion{Fe}{1} 3720 \AA \ absorption line, the relative strengths
of \ion{Ca}{1} and \ion{Ca}{2} lines, and the depth of the imaged \ion{Fe}{2}
absorption spot were used to estimate an iron mass in the form of
\ion{Fe}{2} at between 0.1--1.0 M$_{\odot}$.

Subsequent high-resolution {\sl HST} Advanced Camera for Surveys (ACS) images taken in
2004 showed that the \ion{Ca}{2} absorption is roughly spherical, with a maximum
diameter of $0\farcs 80 \pm 0\farcs 05$ \citep{Fesen2007}. At the estimated
distance of $780 \pm 20 \unit{kpc}$ for M31 \citep{McConn2005,Dalcanton2012},
this angular diameter corresponds to a mean expansion velocity of $12,400
\unit{km} \unit{s}^{-1}$ over the 119 yr age of SN~1885 in 2004. 

The agreement between the remnant's size and its expansion velocity as measured
in the \ion{Ca}{2} absorption implies that the SN~1885 remnant is still largely
in free expansion.  Possible detections of the remnant in radio \citep{SD01}
and X-ray \citep{Hof13} emission have been reported, but the faintness of these
possible detections is consistent with the evidence that the bulk of the ejecta
remain in free expansion.


More recent {\sl HST} images obtained in 2010 and 2012 with the ACS and the
Wide Field Planetary Camera (WFC3) show extended low velocity \ion{Fe}{1}
absorption offset to the east from the remnant's center defined by \ion{Ca}{2}
images \citep{Fesen2015}.  Most striking is the appearance of the remnant in
\ion{Fe}{2} absorption, which appears to be concentrated in four plumes
extending from the remnant's center to at least $10{,}000 \unit{km}
\unit{s}^{-1}$, in sharp contrast to the approximately spherical distribution
of \ion{Ca}{2}.

Because its ejecta are in near-free expansion, the distribution of elements in
the remnant of SN~1885 (hereafter SNR~1885) is essentially the same as that
shortly after the explosion.  The dominant element near the center of the
remnant at the present time is expected to be iron, with Fe${}^+$ the
dominant iron ionization species \citep{Hamilton00}. 

SNR~1885's UV absorption spectrum is predicted to be especially rich between
$2000$ -- $3000 \unit{\AA}$ which contains several resonance lines of
\ion{Fe}{1} and \ion{Fe}{2} \citep{Fesen1999}. However, the M31 bulge is
faint in this wavelength region, thus requiring long exposures to achieve
adequate signal-to-noise.

Here we present optical and UV {\sl HST} spectra of SNR~1885 taken as part of a
spectral campaign in Cycle 21 of {\sl HST\/} observations. The data were
obtained to investigate the velocity distribution of neutral and singly ionized
species of magnesium, calcium, and iron.  The observations are described in \S2
with the results presented in \S3, discussed in \S4, and implications for
SN~Ia presented in \S5.

\begin{deluxetable}{llcr}
\tabletypesize{\footnotesize}
\tablecolumns{4}
\tablewidth{0pt}
\tablecaption{STIS Observations \label{tab:observations}}
\tablehead{\colhead{Detector} &\colhead{Observation Date} &\colhead{Slit Position} &\colhead{Exposure}}
\startdata
CCD & 2013 October 27 & 1B & $10{,}785 \unit{s}$\\
CCD & 2013 October 21 & 2B & $10{,}785 \unit{s}$\\
CCD & 2013 October 26 & 3B & $10{,}785 \unit{s}$\\
CCD & 2013 December 23 & 1A & $10{,}785 \unit{s}$\\
CCD & 2013 December 09 & 2A & $10{,}785 \unit{s}$\\
CCD & 2013 December 10 & 3A & $10{,}785 \unit{s}$\\
& & &\\
MAMA & 2013 December 26 & \nodata & $10{,}491 \unit{s}$\\
MAMA$^1$ & 2013 December 26 & \nodata & $\phn 7{,}701 \unit{s}$\\
MAMA & 2013 December 30 & \nodata & $10{,}491 \unit{s}$ \\
MAMA & 2014 January 02  & \nodata & $10{,}491 \unit{s}$ \\
\enddata
\tablenotetext{1}{Not used in this study.}
\end{deluxetable}

    \begin{figure}
    \begin{center}
    \leavevmode
    \includegraphics[scale=.7]{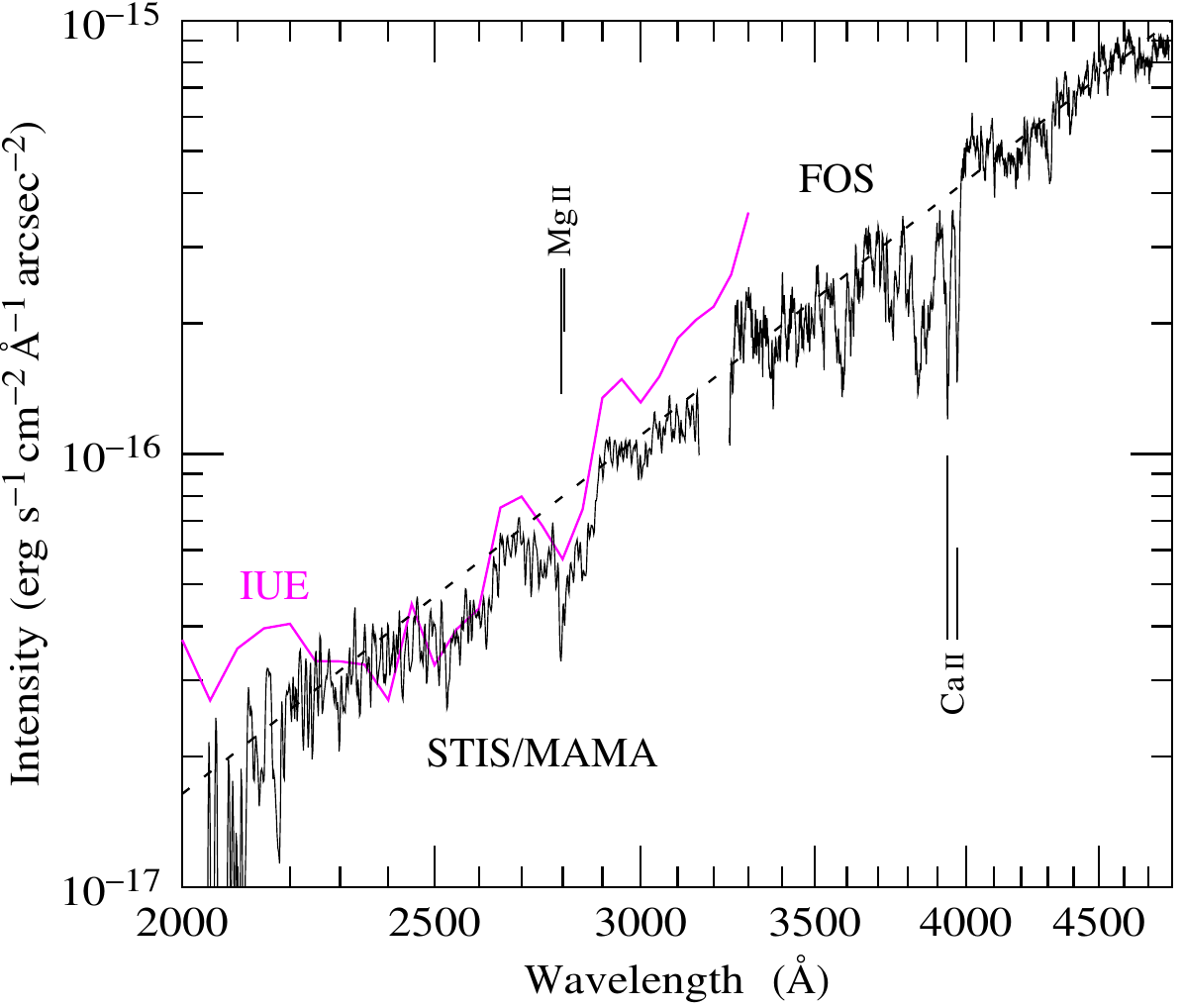}
    \caption{ \label{stismama_ccd_bkgd}
STIS/MAMA rest-frame background spectrum of the bulge of M31
at the location of SNR~1885 (black line). 
Shown for comparison are
the spectrum of the bulge at the position of SNR~1885
taken with the FOS
\citep{Fesen1999} (upper right),
and the spectrum of the nucleus of M31 taken with
IUE through a diameter $14 \unit{arcsec}$ aperture
\citep{Burstein1988} (lower left).
The faint dashed line is a power law with intensity $\propto \lambda^{4.7}$.
    }
    \end{center}
    \end{figure}



\begin{figure*}[hb]
        \centering
         \includegraphics[scale=.75,angle=270]{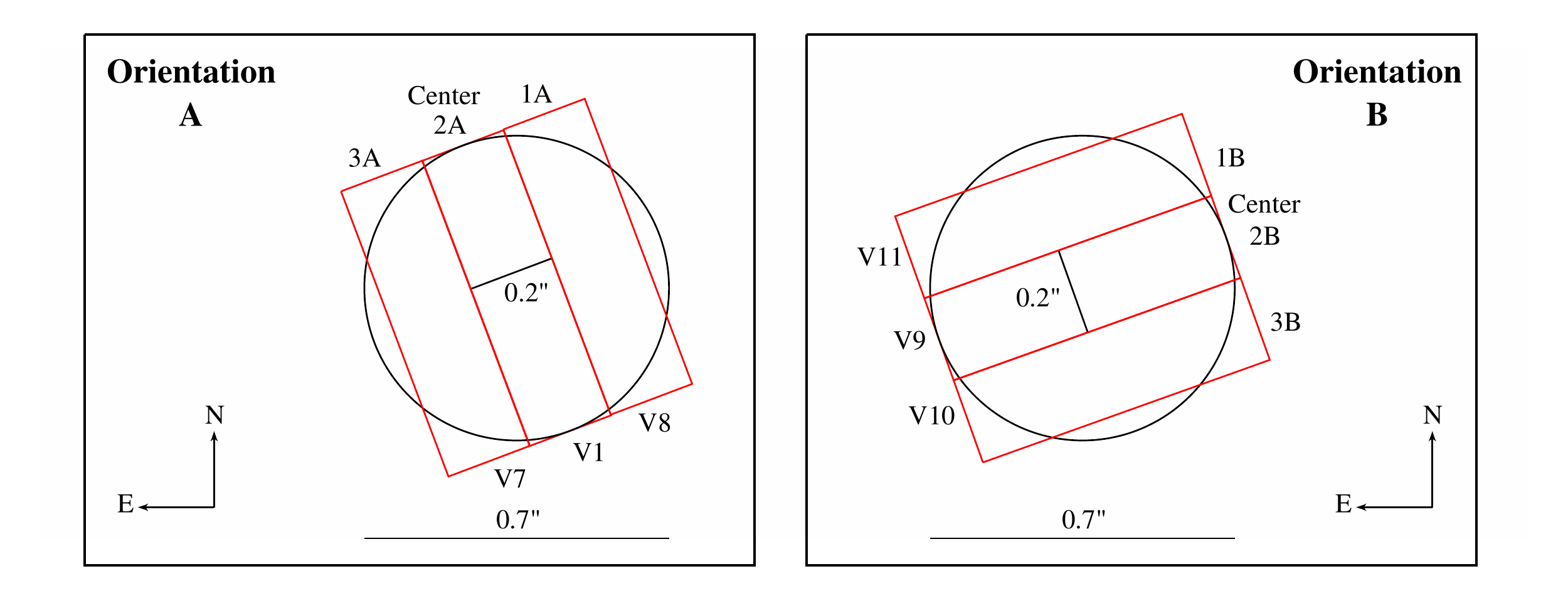}
\caption{
\label{CCD_Slits}
Diagram of the STIS/CCD $0\farcs2 \times 52''$slit positions for the observations of SNR~1885. Three
slit positions were used
in two orientations (A and B).  Actual visit numbers (V1, V7, etc) are
noted. Circles $0\farcs7$ in diameter
correspond to an expansion velocity of 10,000 km s$^{-1}$.
}  
\end{figure*}


\begin{figure*}[t]
\centering \includegraphics[scale=.40]{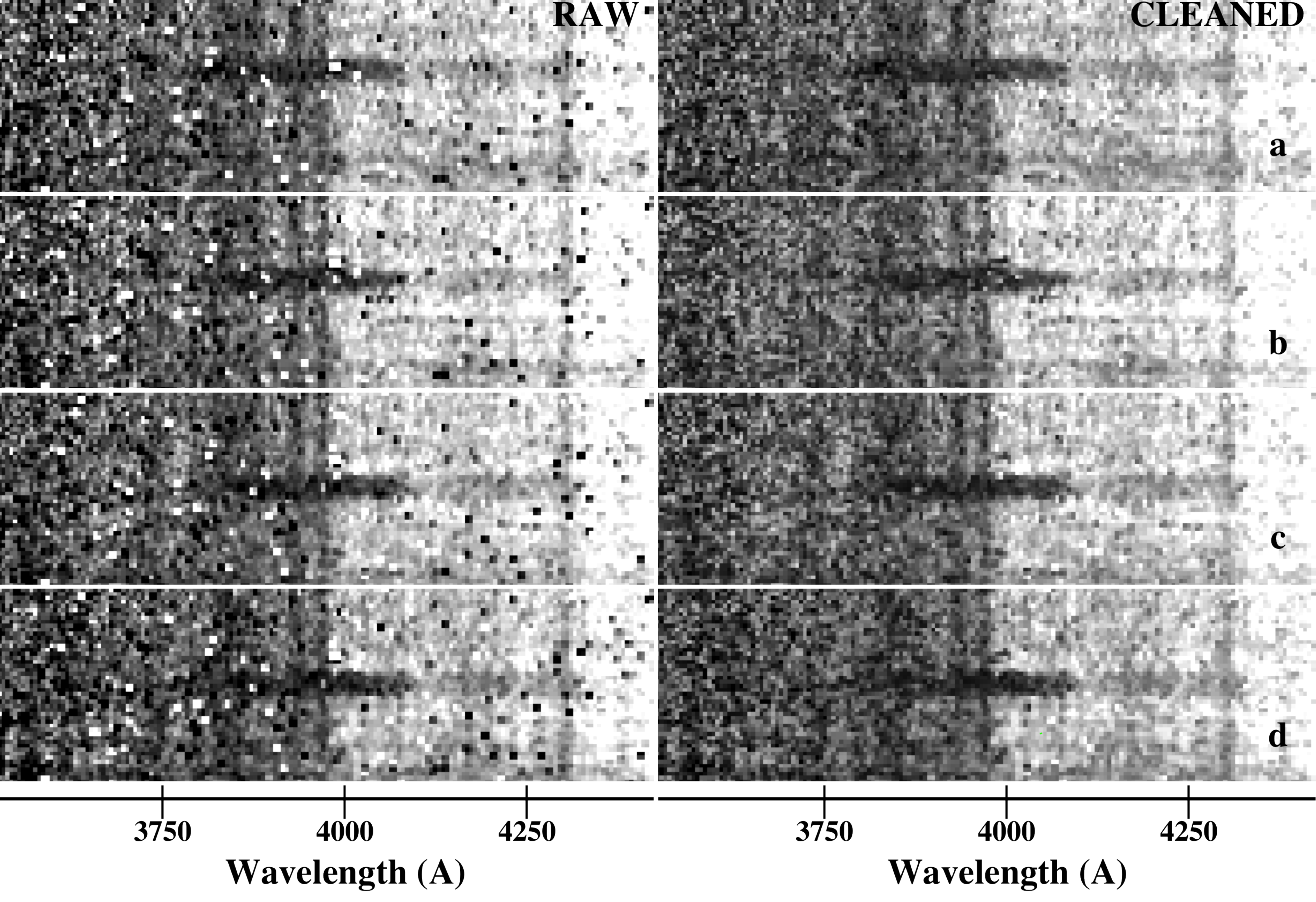}
\caption{The STIS/CCD spectrum of SNR~1885 between $3500\unit{\AA}$ and $4500\unit{\AA}$ for each of the 4 orbits
taken at Slit 2B, illustrating hot and cold pixel corrections. 
Each rectangle show a 5.0 arcsec long region along the slit in the y-direction.  
The dark strip features are 
\ion{Ca}{2}~H~\&~K and \ion{Ca}{1} $4227\unit{\AA}$ absorptions from the expanding remnant. 
\textit{Left}: Raw
individual 2D spectra before pixel corrections with the same CCD section
shown in each image. Dithering of SNR~1885 downward along the slit can be seen as a
shift along the CCD except for orbit `d'.   
\textit{Right}: Same individual spectra
after hot and cold pixel corrections. }
\label{Sample_Raw_Imreplace_Fixpix_EntireVisit}
\end{figure*}


\begin{figure*}[t]
    \begin{center}
   \leavevmode
   \includegraphics[scale=0.35]{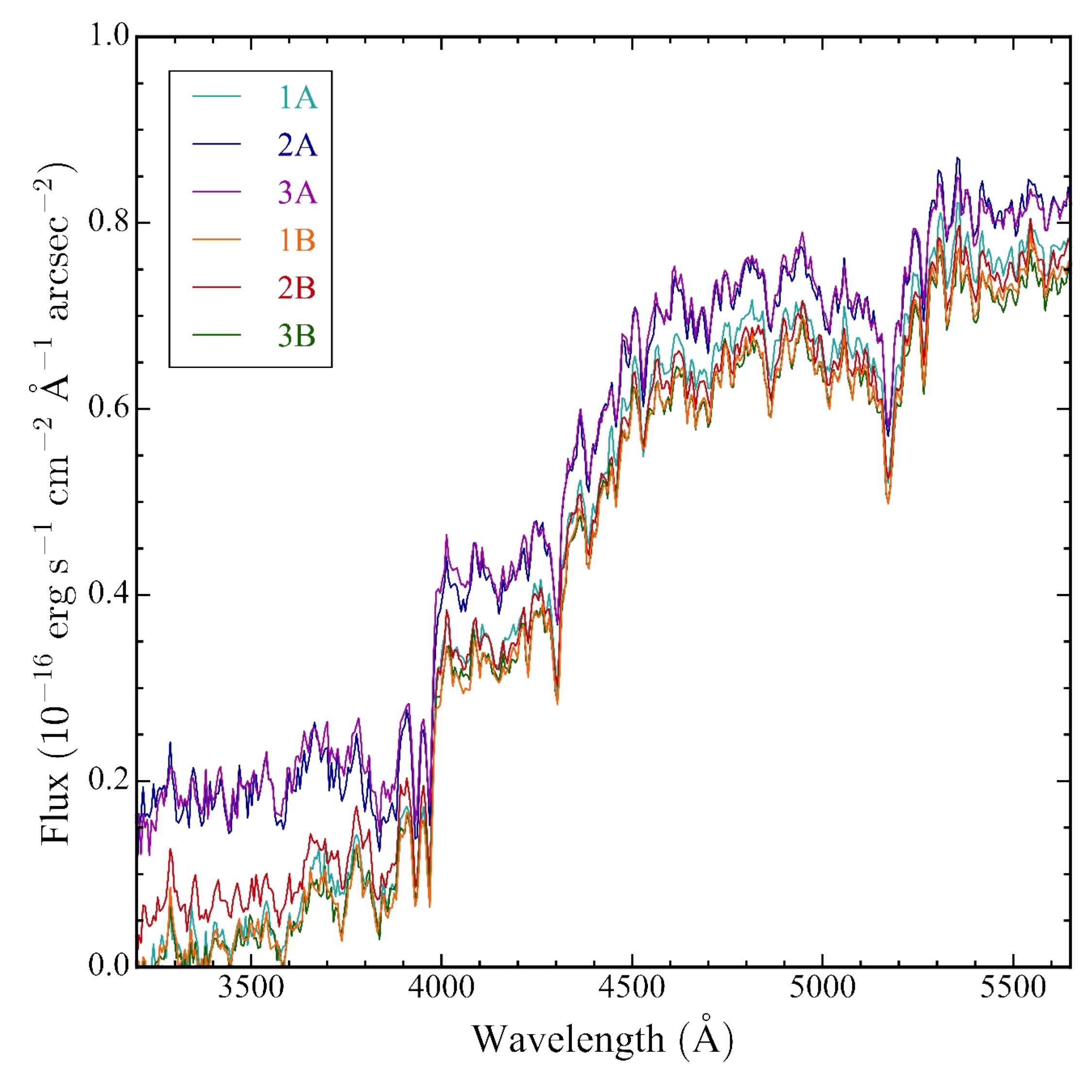}
   \hspace{1em}
   \includegraphics[scale=0.35]{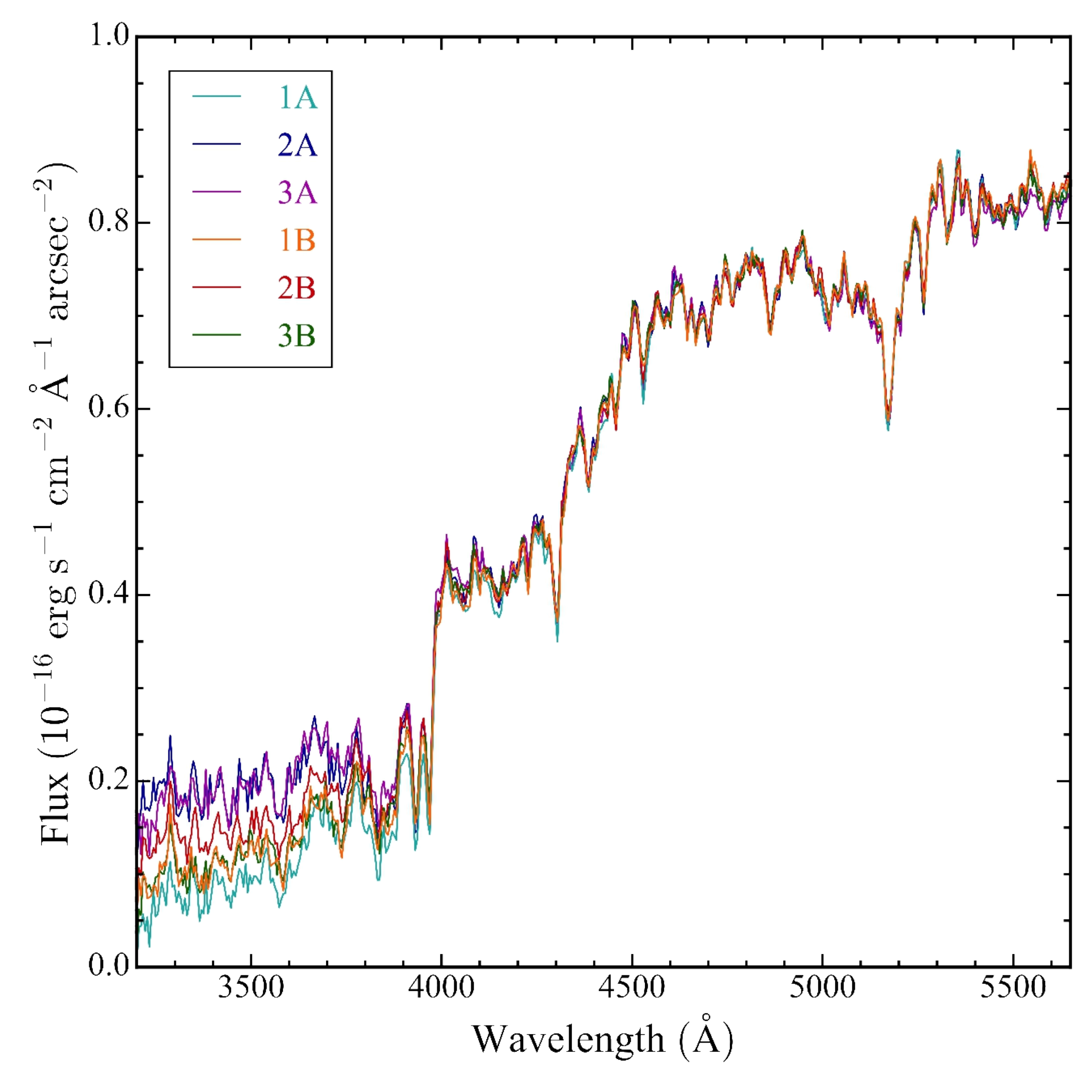}
\caption{ \label{M31_bkcorrection} \textit{Left}:  
STIS/CCD spectra of the bulge of M31 at the location of SNR 1885 
for each of the six visits. The observed background spectra 
show clear systematic differences. 
\textit{Right}: The STIS/CCD background spectra adjusted by a visit-dependent constant offset.}
\end{center}
\end{figure*}


\begin{figure*}[t]
    \begin{center}
   \leavevmode
   \includegraphics[scale=0.42]{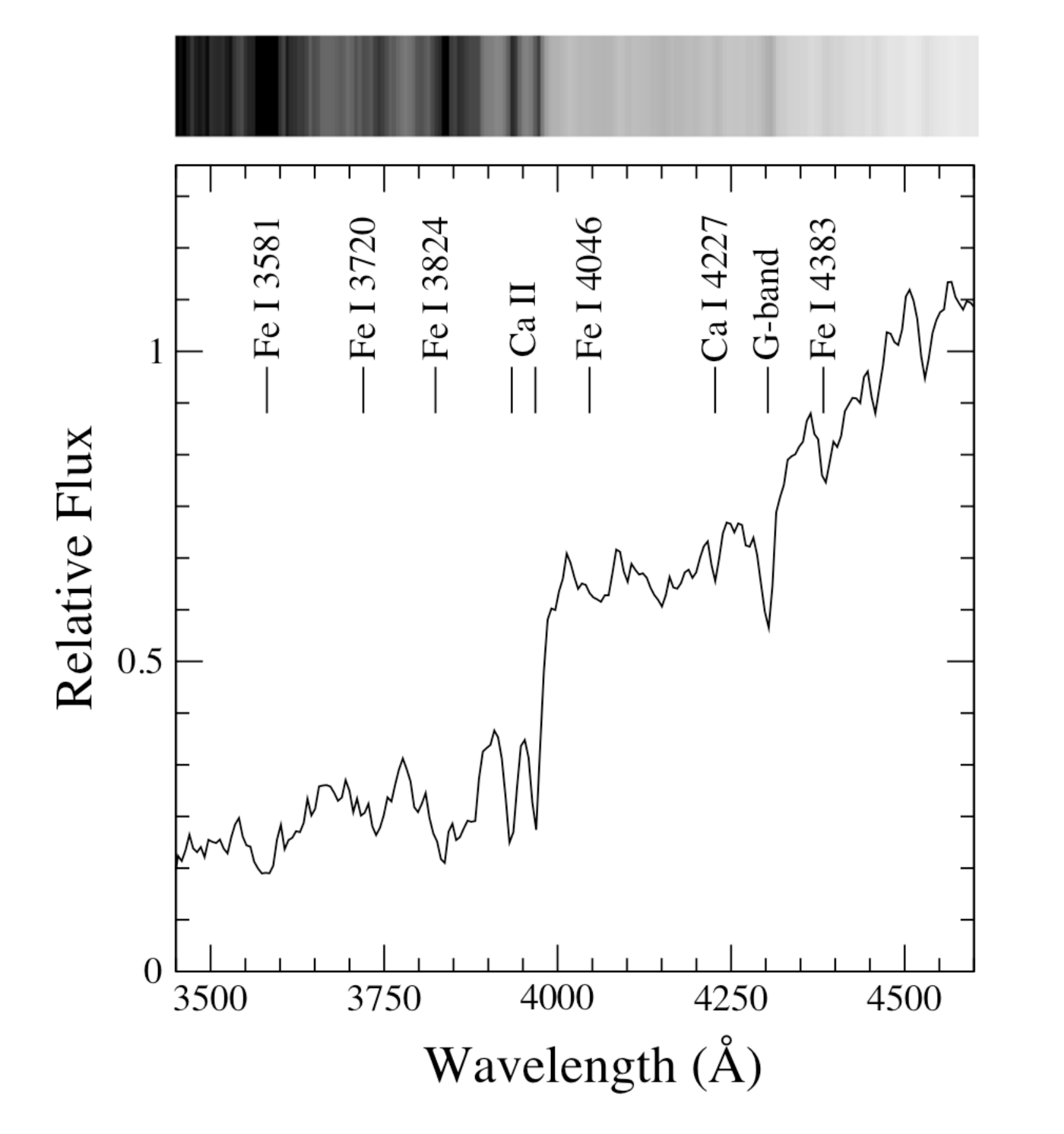}
   \hspace{1em}
   \includegraphics[scale=0.35]{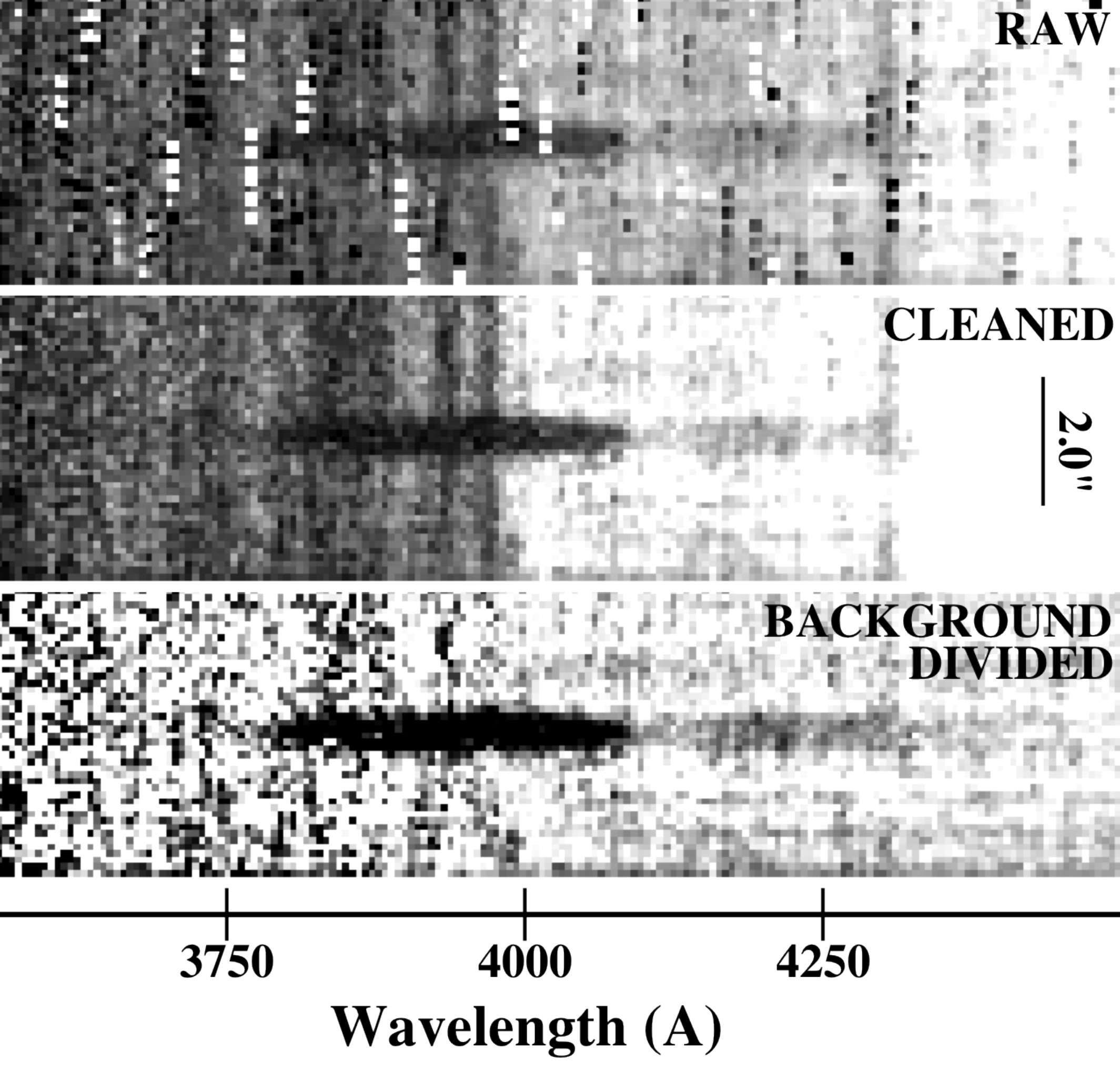}
\caption{ \label{double_figure} \textit{Left}: STIS/CCD spectrum of the M31 bulge extracted
from all slit positions 
with major stellar absorption features labeled. \textit{Right:} Example of data reduction steps:
Raw co-aligned STIS/CCD spectrum 
for slit position 2B (top),
co-aligned and averaged spectrum after hot/cold pixel
corrections (middle), and resulting spectrum following background division by M31's bulge
spectrum (bottom). }
\end{center}
\end{figure*}


\begin{figure*}[t]
        \centering
        \includegraphics[scale=0.4]{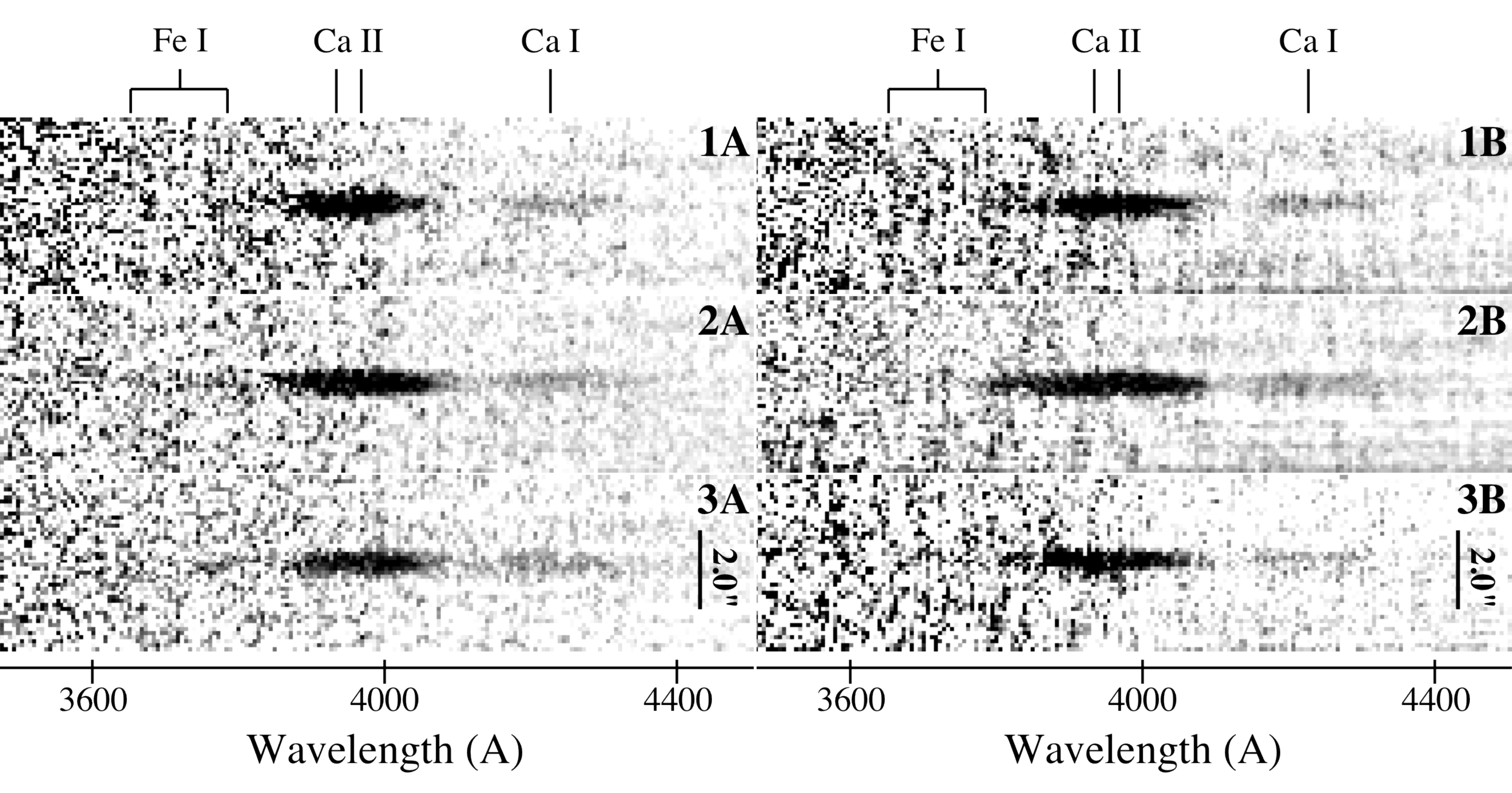} 
\caption{Final 2D background-corrected STIS/CCD spectra for SNR~1885 at each of the six slit
positions shown in a log stretch. Rest wavelengths of \ion{Ca}{2}~H~\&~K, and \ion{Ca}{1}
absorption lines are marked. Locations of  extremely weak red and blueshifted \ion{Fe}{1} 3720 \AA \ 
absorptions are marked but is best seen in the spectra of slit 3A.}
\label{2D_All6Visits}
\end{figure*}


\section{Observations and Data Reduction}


Both optical and UV spectra of SNR~1885 and adjacent M31 bulge were
obtained using the Space Telescope Imaging Spectrograph (STIS) in late 2013 and
early 2014 as shown in Table~\ref{tab:observations}.  Optical spectra were
obtained using the STIS CCD detector (STIS/CCD) which employs a $1024 \times
1024$ pixel CCD covering a $52'' \times 52''$ field-of-view with an imaging
scale of $0\farcs0508$ per pixel.  Because the M31 bulge is faint in the near
UV, on-chip binning was performed with $2 \times 2$ pixel binning, resulting in
a final image scale of $0\farcs1016$ per pixel. STIS/CCD spectra of SNR~1885
used a $52''\times0\farcs2$ slit and the G430L grating covering $2900$ -- $5700$
\AA \ with a spectral resolution of $5.492 \unit{\AA}$ per pixel.

UV spectra of SNR~1885 covering the wavelength range of $2000$ -- $3000$ \AA \
were obtained with the STIS Multi-Anode Microchannel Array (MAMA) detector.  
The MAMA detector has $1024 \times 1024$
pixels covering a $25'' \times 25''$ field-of-view with a pixel scale of
$0\farcs0248$.  The spectra were taken using a $52''\times0\farcs5$ slit and
the G230L grating covering $1576$ -- $3159$ \AA \ at $1.548 \unit{\AA}$ per pixel.

Before discussing in detail the optical and UV spectral observations and data
reduction, Figure~\ref{stismama_ccd_bkgd} shows a comparison of the reduced
STIS/MAMA UV spectral data for the M31 bulge at the location of SNR~1885
compared to previous observations; namely, UV spectra taken with the
International Ultraviolet Explorer (IUE) reported by \citet{Burstein1988}, and
optical spectra of the bulge obtained with the {\sl HST} FOS by
\citet{Fesen1999}.  As can be seen in this figure, our STIS UV spectra of the M31
bulge agrees reasonably well with previous reported data in terms of flux and
gross spectral energy distribution, giving us confidence in our data reduction
procedures.
 
\subsection{Optical Spectra}

SNR~1885 was observed with the STIS/CCD spectrograph during 24 {\sl
HST\/} orbits.  These data were obtained during six visits, with each visit
consisting of four {\sl HST\/} orbits.  At each visit, spectra at a different
slit position across the remnant were taken as illustrated in Figure~\ref{CCD_Slits}.
Slit spectra were obtained at three abutting
$0\farcs2$ locations orientated in one direction (Slits 1A, 2A, \& 3A), and
then in the orthogonal direction (Slits 1B, 2B, \& 3B).

The net signal-to-noise of the STIS/CCD 2D spectrum of the M31 bulge in the
vicinity of SNR~1885 attained in each $10{,}785 \unit{s}$ visit is 12 -- 20 per
binned pixel at $4000$ -- $4500$ \AA. However, due to the steep drop in the
background M31 bulge flux intensity below $4000 \unit{\AA}$, the signal-to-noise 
falls to 2 -- 6 per pixel at $3200$ -- $4000$ \AA.

Due to radiation damage of the STIS/CCD over the course of its lifetime in low
Earth orbit, a significant number of detector pixels are no longer functioning,
recording either an extremely high (`hot pixel') or low (`cold pixel') flux
value. To facilitate removal of these pixels and to mitigate variations in the
sensitivity of individual pixels, the spectrum of SNR~1885 was dithered along
the slit between successive orbits.  The first two orbits of each visit were
taken in the same position, while the next two orbits of each visit were
dithered by respectively 7 or 14 pixels (A visits) and 3 or 6 pixels (B visits)
(see Fig.~\ref{Sample_Raw_Imreplace_Fixpix_EntireVisit}).

To correct for hot and cold pixels that remained despite spectral dithering, we
set a range of acceptable flux values and then generated a mask of bad pixels
that exceeded this range.  Because the flux of the bulge and the sensitivity of
the CCD varied significantly between the red and blue wavelengths ends of the
spectra, acceptable ranges were varied in narrow wavelength bins across each
spectrum.  

The IRAF task IMEXPR was used to create a mask of bad pixels, and
the IRAF task FIXPIX was used to replace bad pixels using linear interpolation
over neighboring pixels.  An example of this procedure for Slit 2B is
shown in Figure~\ref{Sample_Raw_Imreplace_Fixpix_EntireVisit}, where the left
side corresponds to the raw images and the right side corresponds to the
cleaned images after pixel corrections.

Following these corrections, a 1D background spectrum for each visit was 
generated using the spectra immediately surrounding the remnant. 
To do this, we extracted a background bulge
spectrum using 11 nine pixel wide regions both above and below the remnant for each 
orbit in a given visit. The individual 1D spectra from each of the regions were 
averaged together to form a single 1D spectrum for each visit (see left panel
of Figure~\ref{M31_bkcorrection}).

As Figure~\ref{M31_bkcorrection} shows, we found that there are systematic
large-scale variations between the STIS/CCD spectra for different visits.
Empirically, the differences appear to consist primarily of a constant offset
$c_i$ in the intensity from each visit.  We corrected for these offsets by
adjusting the STIS/CCD background spectra by $\textrm{back}_i=\textrm{off}_i+c_i$, 
where the six constant offsets $c_i$ were calculated such that the mean intensity 
over $4775$ -- $5120\unit{\AA}$ for each visit were in agreement.


The results of these corrections are shown in the right panel of
Figure~\ref{M31_bkcorrection}. As shown in the figure, following these flux
corrections, the background M31 spectra are consistent with each other at
wavelengths above $4000\unit{\AA}$.  The left panel of
Figure~\ref{double_figure} shows the average 1D M31 bulge spectrum with major
stellar absorption features marked.

Next, the individual orbit images for each visit were co-aligned and
averaged to form a single 2D spectrum at each slit position.  The 2D spectrum
was then flux scaled by the same constant offset as the background
for the given visit. We then divided the 2D spectrum of SNR~1885 by the 1D
background spectrum.  

However, as shown in Figure~\ref{M31_bkcorrection}, there were still residual
background spectral differences between visits below 3900 \AA. These
differences are likely due to an uneven stellar distribution in the near UV.
To correct for these differences, we highly smoothed the
residual M31 background spectra from above and below the remnant and
then divided this into the remnant's spectrum for each visit.

The right hand panel in Figure~\ref{double_figure} shows an example of the
three major stages of data reduction for slit position 2B. The final
reduced STIS/CCD spectra for the six individual slit positions are shown in
Figure~\ref{2D_All6Visits}.  

    \begin{figure}
    \begin{center}
    \leavevmode
\includegraphics[scale=.45]{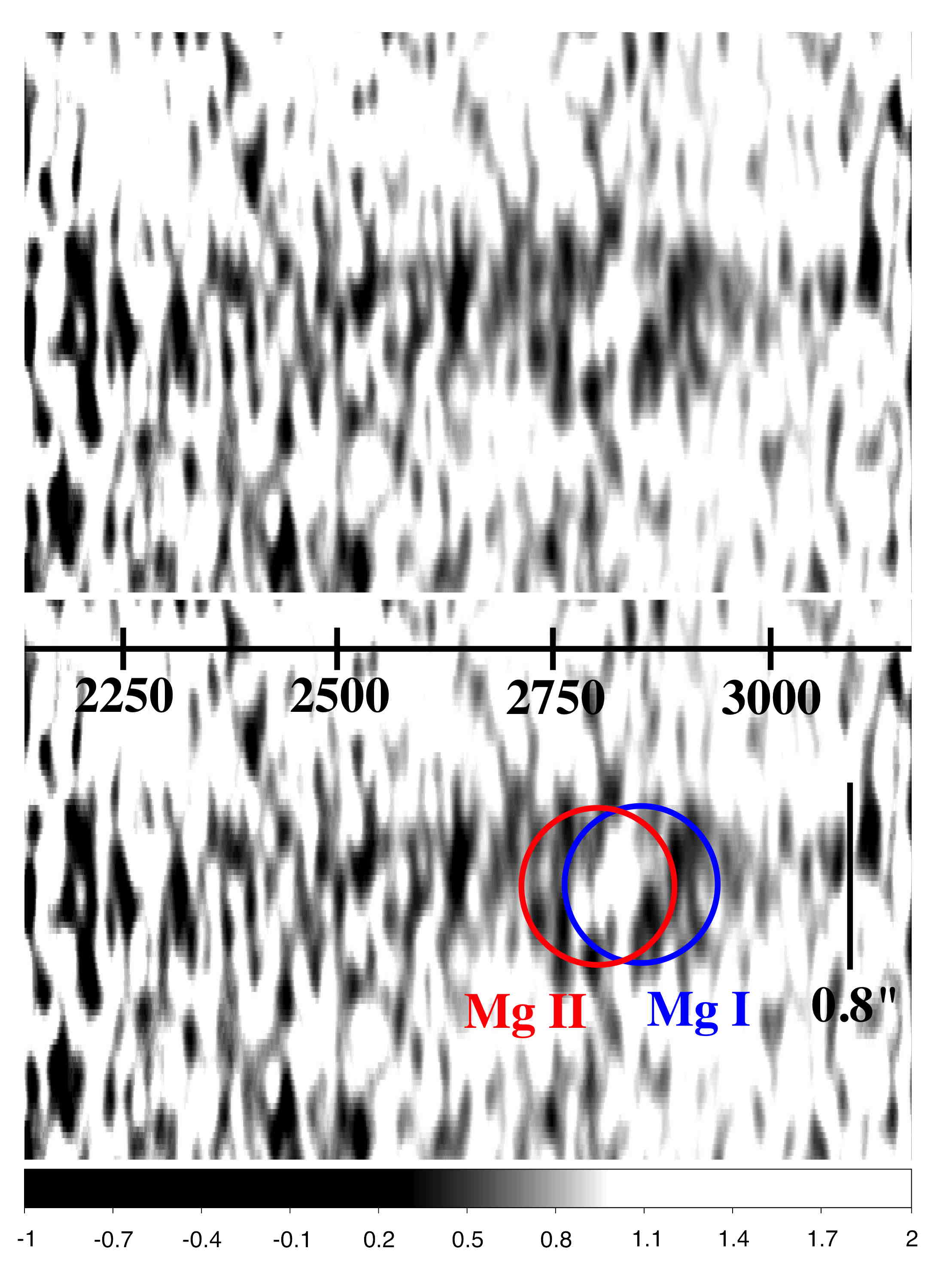}
 \caption{
 \label{stismama2D_BnW}
\textit{Top:} Smoothed, linear stretched, and
background corrected 2D STIS/MAMA spectrum of SNR~1885 covering 2200 to 3100 \AA \
showing a horizontal band of absorption we associate with SNR~1885.
\textit{Bottom:} Same spectrum showing red and blue circles 
centered on \ion{Mg}{2} 2796, 2803
\AA \ and \ion{Mg}{1} 2852 \AA \ suggestive of
a shell of Mg-rich ejecta. The circles have 
diameters of of $1\farcs2$ corresponding to 
average expansion velocity of 9500 km s$^{-1}$ since 1885.
}
    \end{center}
    \end{figure}

    \begin{figure*}
    \begin{center}
    \leavevmode
    \includegraphics[scale=.6]{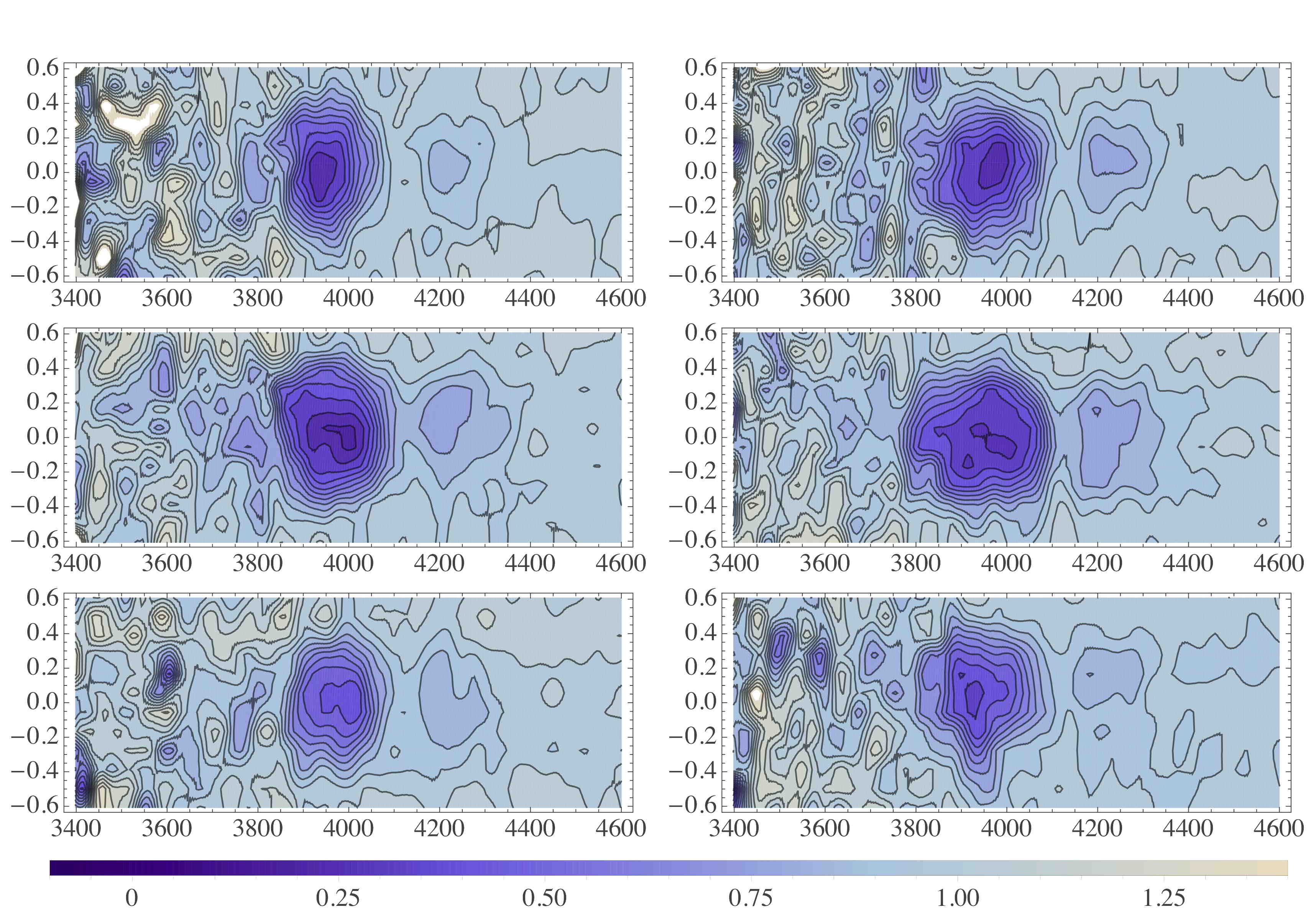}
    \caption{
    \label{2D_blue}
Contour plots of relative observed flux for background-corrected, smoothed STIS/CCD spectra
as a function of wavelength and angular offset
along the slit from the \ion{Ca}{2} center of SNR~1885.
From left to right and top to bottom,
the plots are 1A, 1B, 2A, 2B, 3A, 3B.
Contours are drawn every $0.075$ units of relative flux.
    }
    \end{center}
    \end{figure*}


\subsection{UV Spectra}
\label{stismama-sec}

UV spectra of SNR~1885 were obtained with the STIS/MAMA in 15 orbits split into 
four visits. Three of the four visits consisted of four orbits, while the
remaining visit was only three orbits.  All spectra were taken at a single slit
position with the slit orientation the same as the A configuration of the
STIS/CCD observations (see Fig.\ \ref{CCD_Slits}).  The placement of SNR~1885
during successive orbits within a visit was dithered by $0\farcs3$ along the
slit.

The three four-orbit visits yielded data consistent with each other, but the
shorter three-orbit visit yielded negative mean fluxes suggestive of a calibration
problem; consequently, we discarded those data.  The total exposure time used
in our study was $31{,}473\unit{s}$.  We note that during each visit, the flux
from later orbits was systematically lower than that from earlier orbits, with the
mean flux becoming negative at shorter wavelengths.  Moreover, measured fluxes
were found to be negative below $2000 \unit{\AA}$ and thus untrustworthy below
about $2200 \unit{\AA}$.  

As was done for the STIS/CCD data, M31's bulge spectrum was found using the spectra
surrounding the remnant in the coaligned, summed 2D spectrum. Eleven $1\farcs22$ regions
were extracted from above and below the remnant and averaged together.  The
individual pixels of this 2D image were then averaged to form a single 1D bulge
spectrum for M31.  The spectrum of SNR~1885 was then divided by M31's bulge
spectrum.

The resulting M31 bulge background-corrected UV spectrum of SNR~1885 is shown
in Figure~\ref{stismama2D_BnW}.  The spectrum has been stretched vertically by
a factor of 4.38 so that a spherically symmetric absorbing shell
appears spherical.  Although there is a small uncertainty in the precise
positioning of the slit with respect to the center of SNR~1885, we attribute the
horizontal band of absorption near the nominal slit center to be due to absorption from
SNR~1885.

\begin{figure*}[t]
\centering
\includegraphics[scale=0.35]{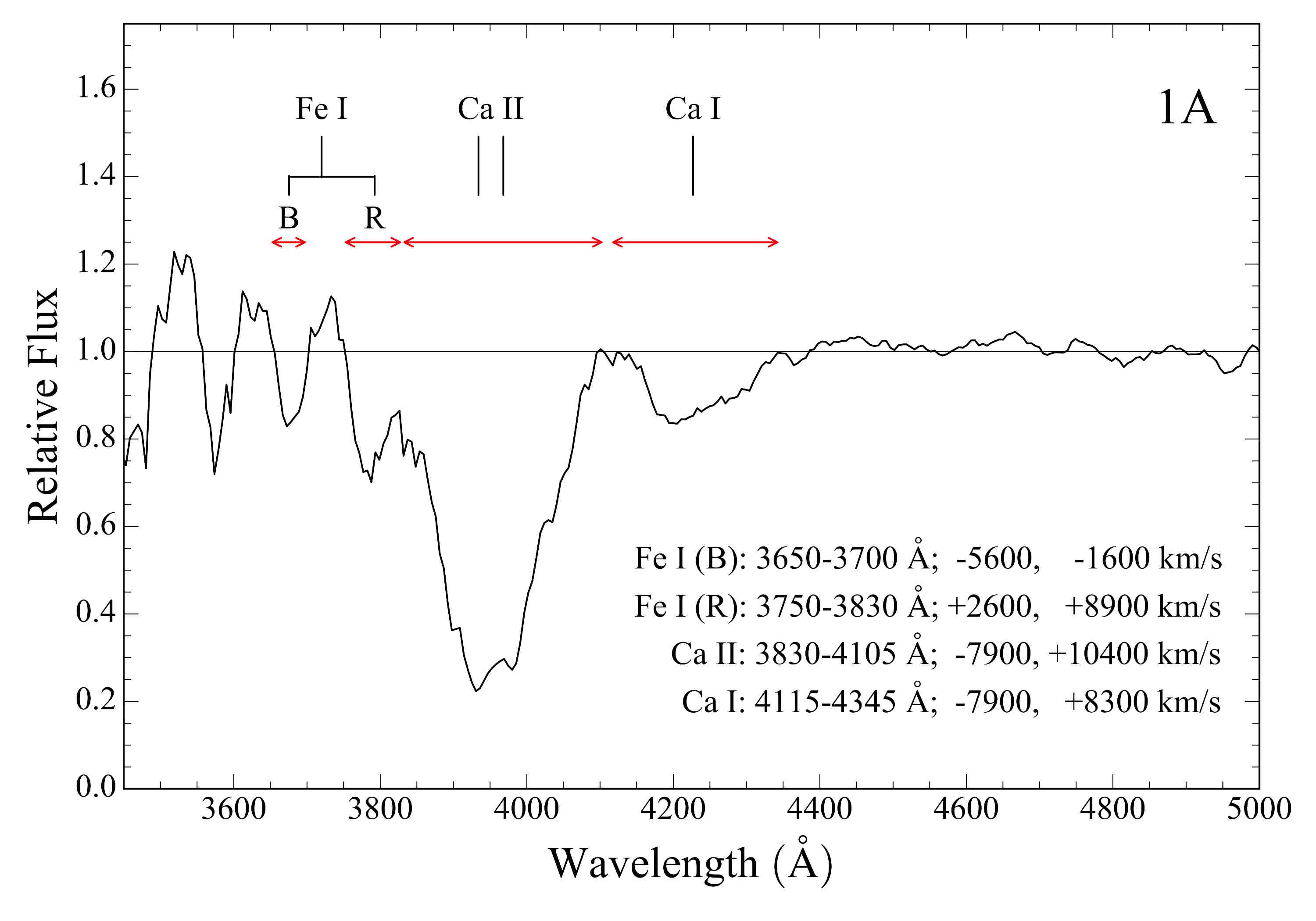}
\includegraphics[scale=0.35]{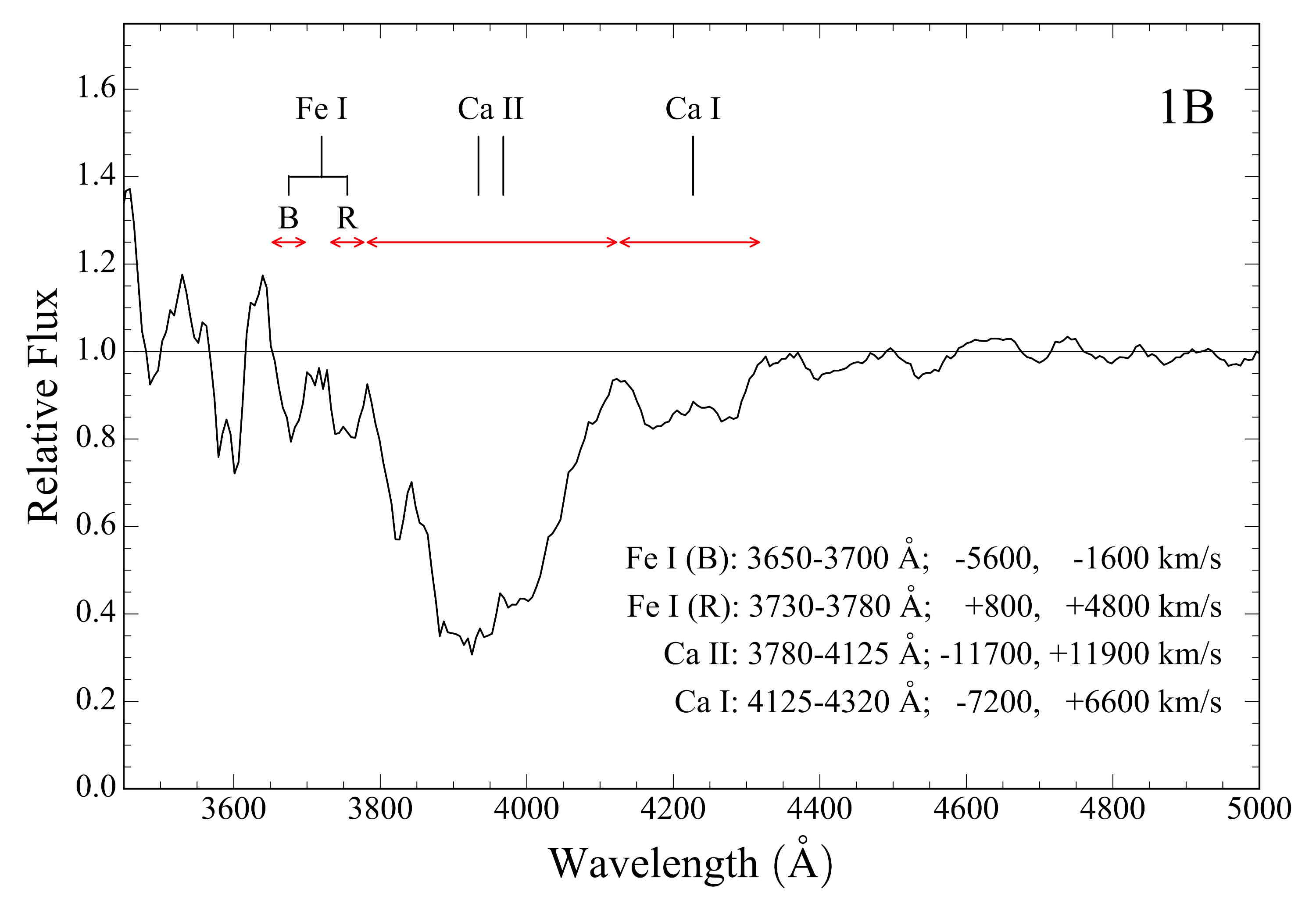}\\
\includegraphics[scale=0.35]{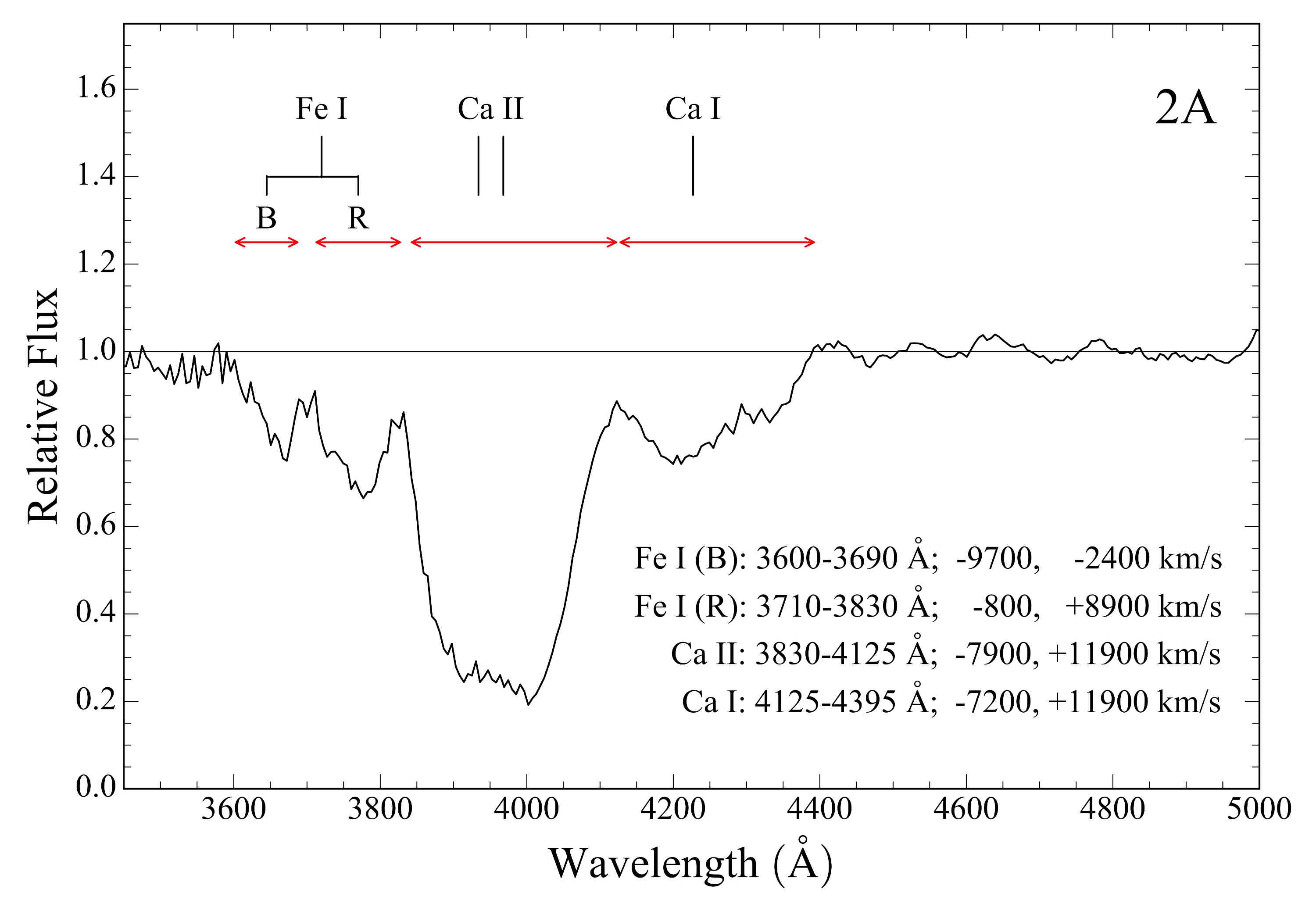}
\includegraphics[scale=0.35]{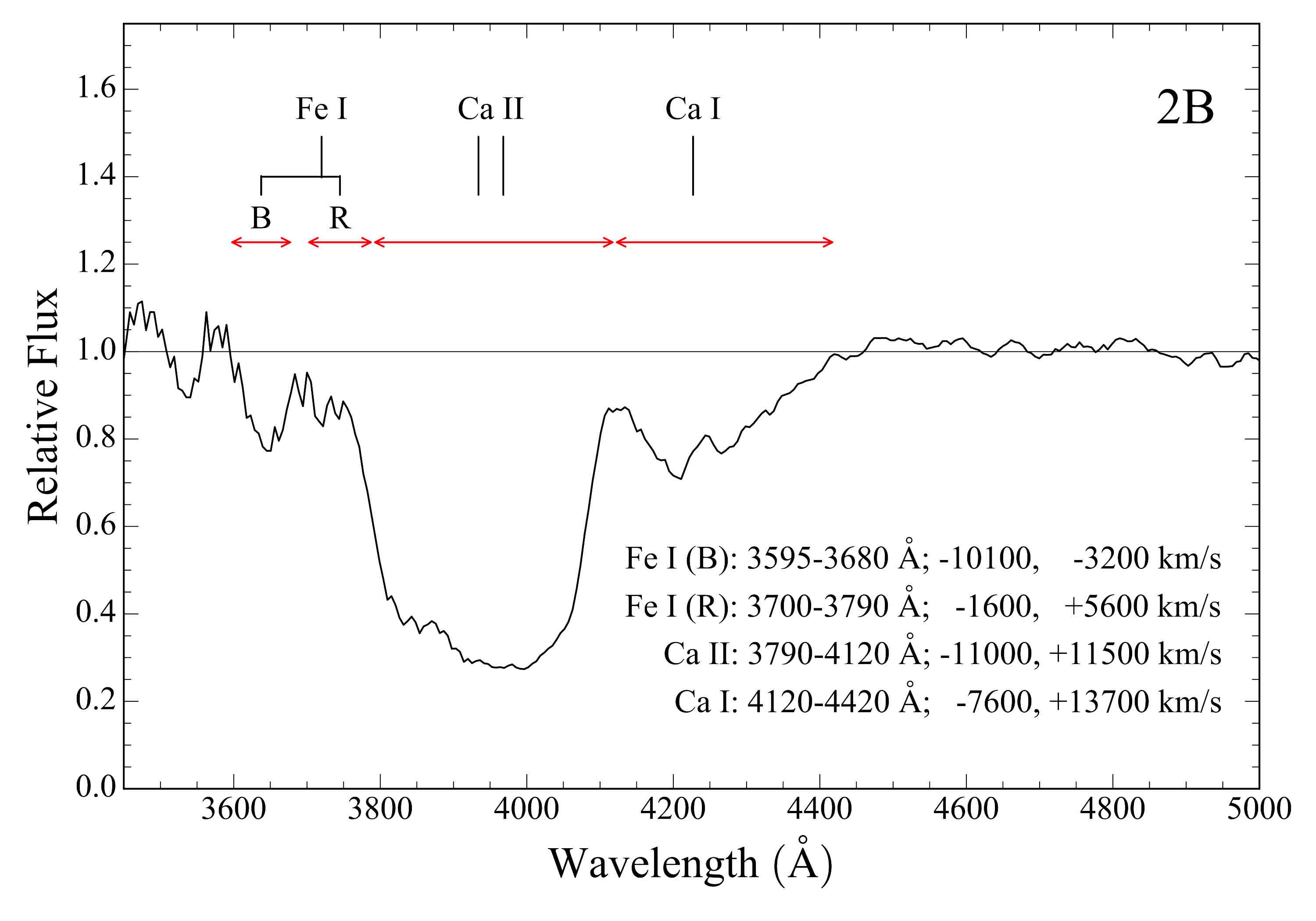}\\
\includegraphics[scale=0.35]{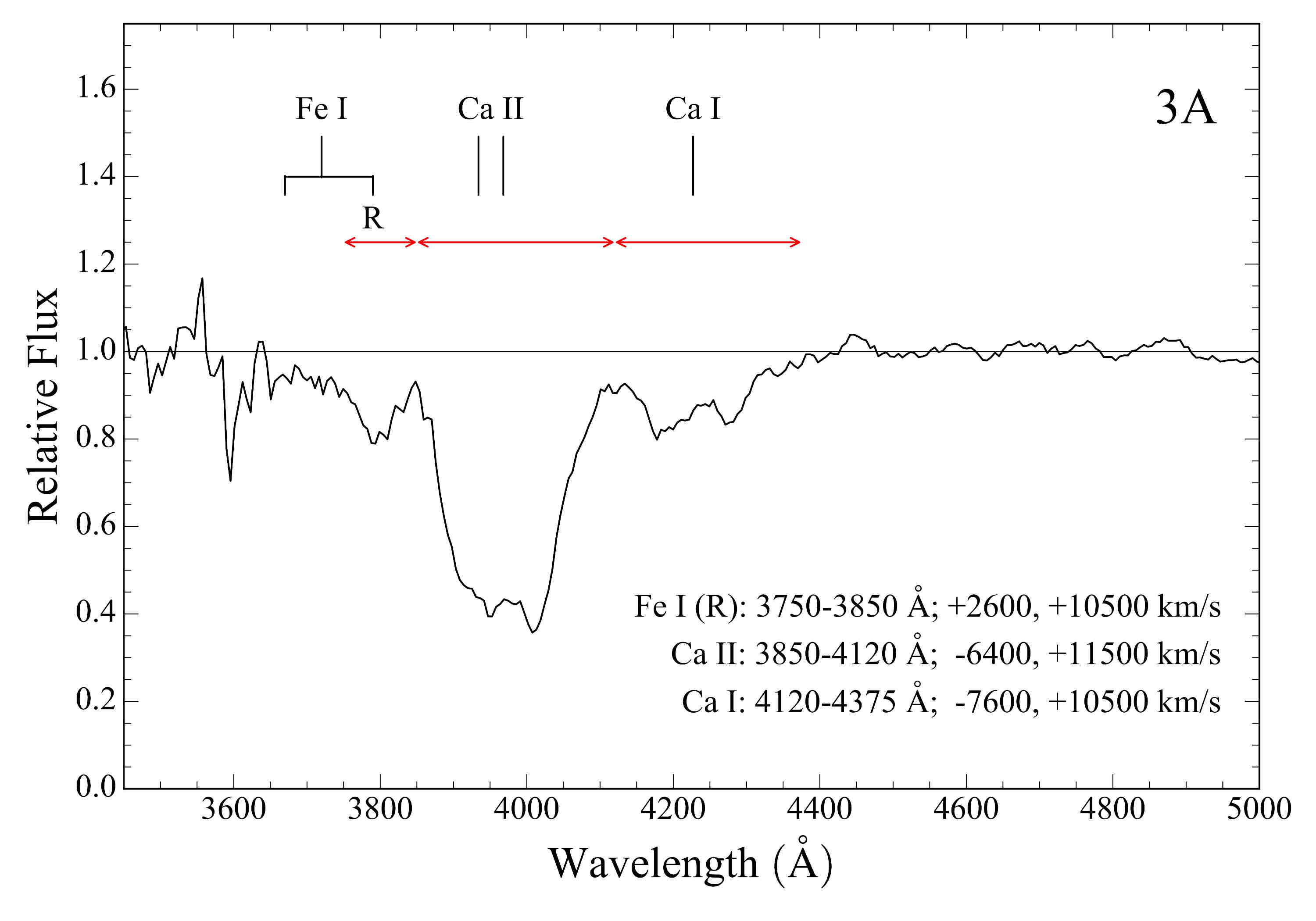}
\includegraphics[scale=0.35]{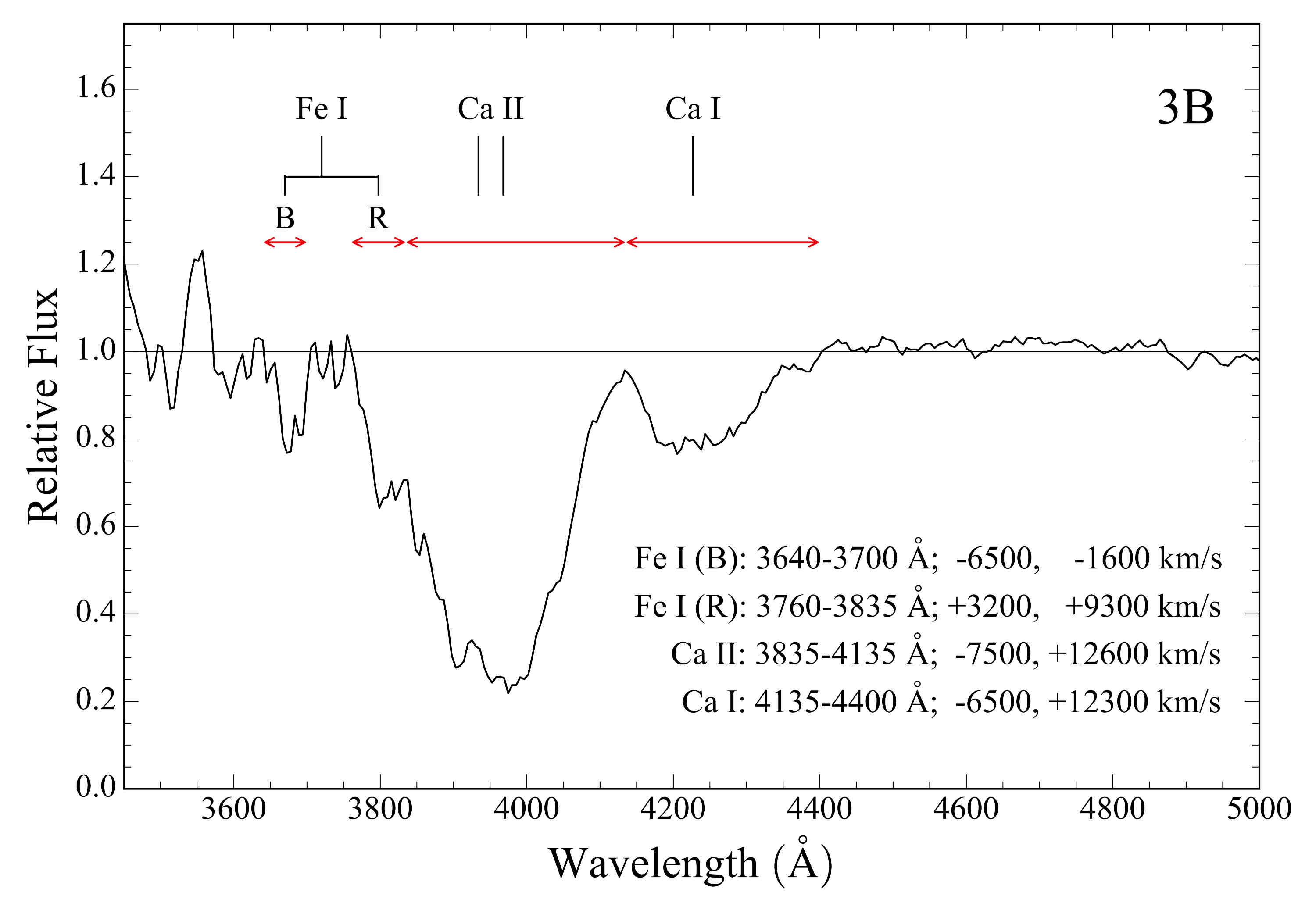}

\caption{The STIS/CCD spectrum for the central $0\farcs3$ for the six slit positions.
Rest wavelengths of \ion{Fe}{1}, \ion{Ca}{2}~H~\&~K, and \ion{Ca}{1}
absorptions are labeled. For \ion{Fe}{1} 3720 \AA, both blueshifted and redshifted absorption features
are seen and indicated by the lines on either side of $3720\unit{\AA}$. 
Red arrows   
mark the approximate widths of the absorption features, with  
wavelengths and velocities  
listed in the bottom right. 
   }
\label{1DSpectra_all_positions}
\end{figure*}

    \begin{figure*}
    \begin{center}
    \leavevmode
    \includegraphics[scale=.4]{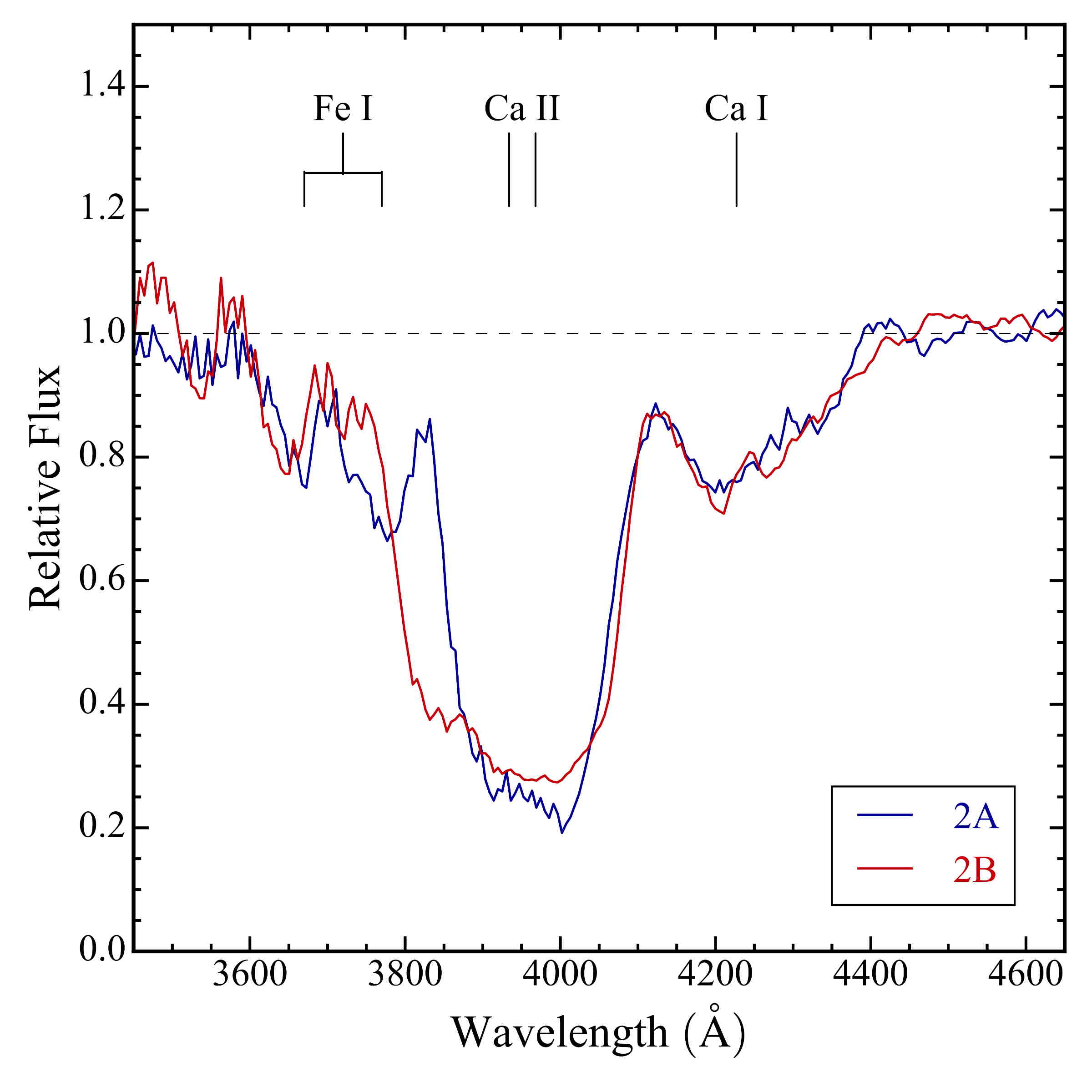}
    \hspace{2em}
     \includegraphics[scale=.4]{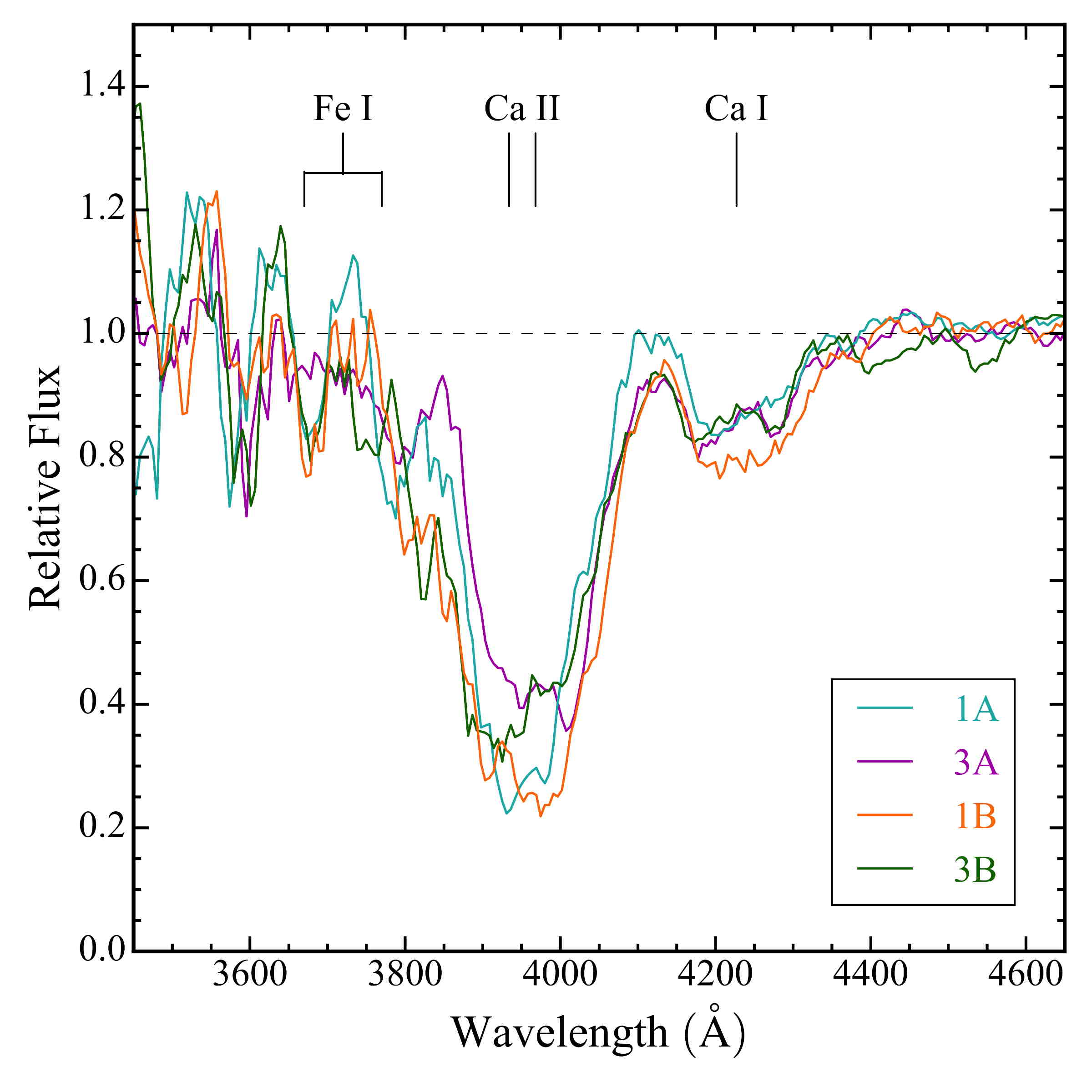}
    \caption{
\label{stisccd_pt1} Comparison of STIS/CCD spectra 
for the central 3 pixels $(0\farcs3)$ of the slit for the two central slits (2A and 2B; left panel) and 
four edge slit spectra (1A, 3A, 1B, and 3B; right panel). 
    }
    \end{center}
    \end{figure*}


    \begin{figure*}
    \begin{center}
\includegraphics[scale=.55]{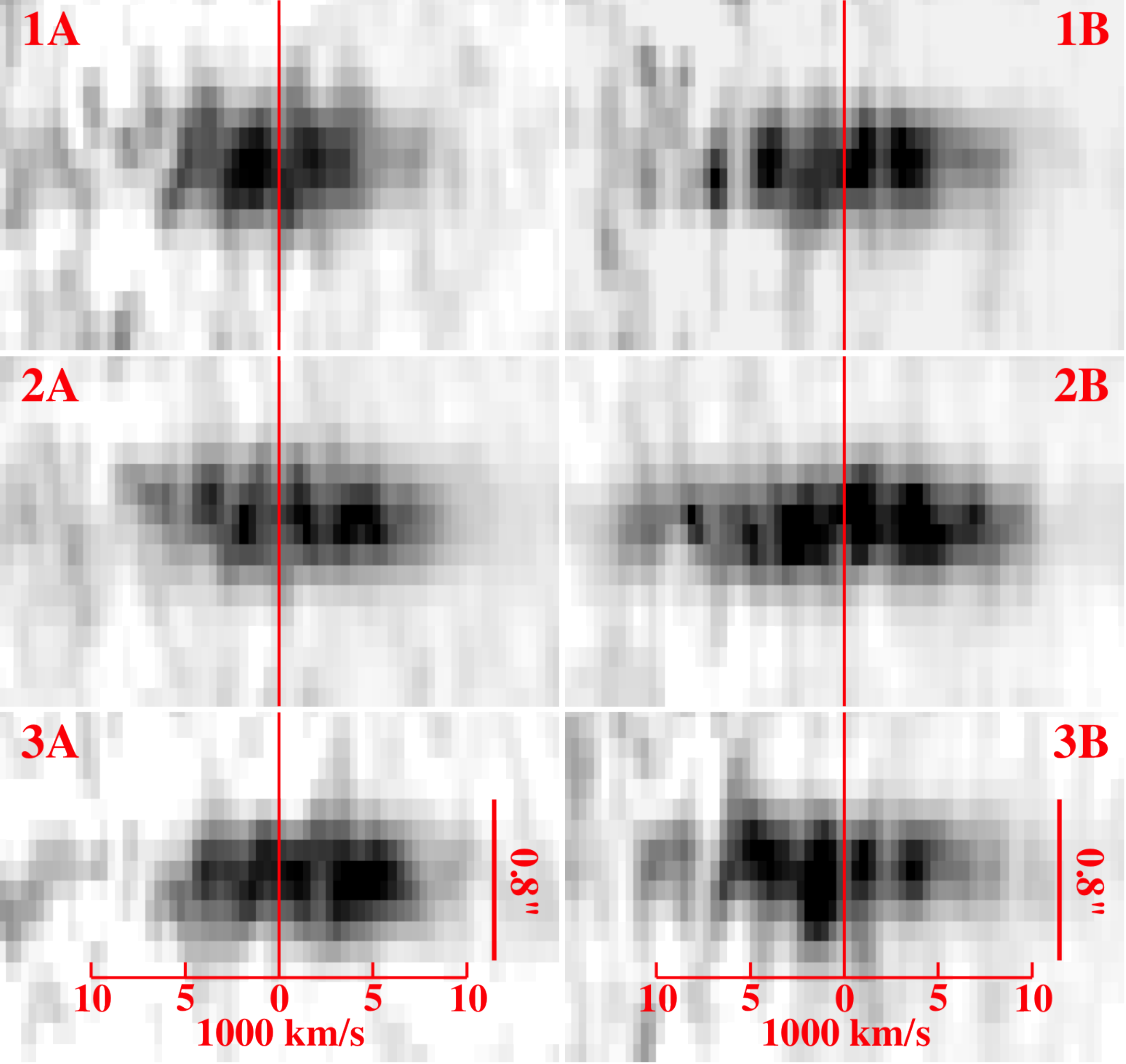}
 \caption{
 \label{Ca_clumps}
Reduced 2D STIS/CCD spectra of the \ion{Ca}{2} absorptions at the six slit
positions, shown in a log stretch, indicating the presence of Ca-rich ejecta clumps.
The vertical line lies at 3950 \AA, marking the approximate
center of the saturated \ion{Ca}{2} absorption.
}
    \end{center}
    \end{figure*}


\begin{deluxetable*}{clrlrcclr}
\scriptsize
\tablecaption{Velocities of Optical SNR~1885 Absorption Features }
\label{widthspectra}
\tablehead{
\colhead{Slit}&
\multicolumn{2}{c}{\underline{~~~~~~~~~~~~~ \ion{Ca}{2} ~~~~~~~~~~~~~~~~~}}&
\multicolumn{2}{c}{\underline{~~~~~~~~~~~~~~ \ion{Ca}{1} ~~~~~~~~~~~~~~~~}}&
\multicolumn{2}{c}{\underline{~~~~~~~~~~~~~ \ion{Fe}{1} (Blue)~~~~~~~~~}}&
\multicolumn{2}{c}{\underline{~~~~~~~~~~~~~ \ion{Fe}{1} (Red)~~~~~~~~~~}}\\
\colhead{Position}&
\colhead{Wavelength}&
\colhead{~Velocity}&
\colhead{Wavelength}&
\colhead{Velocity}&
\colhead{Wavelength}&
\colhead{Velocity}&
\colhead{Wavelength}&
\colhead{Velocity}\\
&
\colhead{$\unit{(\AA)}$}&
\colhead{~$\unit{(km}\unit{s}^{-1})$}&
\colhead{$\unit{(\AA)}$}&
\colhead{$\unit{(km}\unit{s}^{-1})$}&
\colhead{$\unit{(\AA)}$}&
\colhead{$\unit{(km}\unit{s}^{-1})$}&
\colhead{$\unit{(\AA)}$}&
\colhead{$\unit{(km}\unit{s}^{-1})$}\\
\colhead{(1)}&
\colhead{(2)}&
\colhead{~(3)}&
\colhead{(4)}&
\colhead{(5)}&
\colhead{(6)}&
\colhead{(7)}&
\colhead{(8)}&
\colhead{(9)}
}
\tablecolumns{9}
\startdata
1A   & $3830{-}4105$ &$-7900$, $+10400$& $4115{-}4345$ &$-7900$, \phn$+8400$& $3650{-}3700$ &\phn$-5600$, $-1600$& $3750{-}3830$& $+2600$, \phn$+8900$\\
2A   & $3830{-}4125$ &$-7900$, $+11900$& $4125{-}4395$ &$-7200$, $+11900$& $3600{-}3690$ &\phn$-9700$, $-2400$& $3700{-}3830$& $-800$, \phn$+8900$\\
3A   & $3850{-}4120$ &$-6400$, $+11500$& $4120{-}4375$ &$-7600$, $+10500$& \nodata       &\phn \nodata        & $3750{-}3850$& $+2600$, $+10500$\\
1B   & $3835{-}4135$ &$-7500$, $+12600$& $4135{-}4400$ &$-6500$, $+12300$& $3640{-}3700$ &\phn$-6500$, $-1600$& $3760{-}3835$& $+3200$, \phn$+9300$\\
2B   & $3790{-}4120$ &$-11000$, $+11500$& $4120{-}4420$ &$-7600$, $+13700$& $3595{-}3680$ &$-10100$, $-3200$& $3700{-}3790$& $-1600$, \phn$+5600$\\
3B   & $3780{-}4125$ &$-11700$, $+11900$& $4125{-}4320$ &$-7200$, \phn$+6600$& $3650{-}3700$ &\phn$-5600$, $-1600$& $3730{-}3780$ & ~ $+800$, \phn $+4800$ \\
\enddata
\tablecomments{ Columns:
(1) -- Slit positions.
(2 \& 3) -- Wavelength and velocity of \ion{Ca}{2} absorption features.
Note: Velocities are underestimates
due to the overlap between the \ion{Ca}{1} $4227\unit{\AA}$ and \ion{Fe}{1} $3720\unit{\AA}$ absorption features.
(4 \& 5) -- Wavelength and velocity of the \ion{Ca}{1} $4227\unit{\AA}$ absorption features.
The blueshifted  velocities are underestimates
due to overlap with redshifted \ion{Ca}{2} absorption. The red wavelengths should be a truer
measure of the maximum radial velocity.
(6 \& 7) -- Wavelength and velocity of the blue side of the shell of \ion{Fe}{1} $3720\unit{\AA}$.
(8 \& 9) -- Wavelength and velocity of the red side of the shell
of \ion{Fe}{1} $3720\unit{\AA}$. The wavelengths and velocities on the red side of this feature
may be underestimates due to the overlap with blueshifted \ion{Ca}{2} absorption.
}
\end{deluxetable*}




\section{Results}

The goal of taking these STIS spectra was to investigate the three dimensional
structure of Ca, Fe, and Mg rich ejecta in SNR~1885.  For example, variations
in \ion{Ca}{2} absorption might indicate regions of increased or decreased
Ca$^{+}$ concentrations much like those seen in  narrow-passband {\sl HST}
\ion{Ca}{2} images \citep{Fesen2007,Fesen2015}.  These spectra also allow a
comparison  with a 1996 FOS spectrum of SNR~1885  \citep{Fesen1999} to look for
changes in absorption in the intervening 17 year period. 

Below we describe the observed optical and UV absorption features for SNR~1885,
limited by the signal-to-noise for the weaker  \ion{Ca}{1} and \ion{Fe}{1}
features and the highly saturated nature of the \ion{Ca}{2} H \& K line
absorption blend.  In addition, due to the high velocities observed in the
remnant, there is significant overlap between these absorption features, which
limits measurements of the maximum velocities observed for \ion{Ca}{2},
the maximum blueshifted velocity for \ion{Ca}{1}, and the maximum
redshifted velocity for \ion{Fe}{1}.

\subsection{\ion{Ca}{2}}

While there are variations in the remnant's \ion{Ca}{2} absorption
among the slit spectra, they indicate a fairly spherical remnant.
Figure~\ref{2D_blue} shows 2D intensity contour plots of the six
background-corrected STIS/CCD spectra covering the remnant's \ion{Fe}{1} 3720
\AA, \ion{Ca}{2} 3934, 3968 \AA, and \ion{Ca}{1} 4227 \AA \ resonance line
absorptions.  The contours have been stretched vertically by a factor of seven
so that a spherical distribution of absorbing ejecta appears spherical in
these images.  The spectra have also been smoothed with a Gaussian with
1-$\sigma$ width of 3 pixels in the horizontal (wavelength) direction, and 1/2
pixel in the vertical (angle along the slit) direction, so that the image
resolution is comparable in both directions.
As expected, the two central spectra, Slits 2A and 2B, appear the most
spherical, with the \ion{Ca}{2} absorption at Slit 2B affected by some
redshifted \ion{Fe}{1} 3720 \AA \  absorption.  

The overlap of the remnant's broad \ion{Ca}{1} and \ion{Ca}{2} absorptions is
illustrated in Figure~\ref{1DSpectra_all_positions} where we present smoothed
1D spectral plots of all six slit positions.  The plots were made using
background-corrected spectra averaged over the central 3 pixels
($0\farcs3$) since they contain the highest signal-to-noise data and avoid
contamination from the surrounding M31 background.  In these plots, we mark
rough estimates for the maximum and minimum wavelengths for \ion{Ca}{1},
\ion{Ca}{2}, blueshifted and redshifted \ion{Fe}{1} absorptions along with implied
velocities (cf.\ Table 2).

The left panel of Figure~\ref{stisccd_pt1} shows a comparison of the 1D spectra 
of SNR~1885 for the two center slit positions, while the right panel shows a 
comparison of the spectra from the edge slit positions.  The depth of the \ion{Ca}{2}
absorption is approximately equal for the central slits, and similar 
but somewhat weaker for the four edge slits, which is consistent 
with a spherically symmetric explosion. However, the detailed shapes of 
the \ion{Ca}{2} absorption feature differ between slits.

Differences between the two central slit spectra (Slits 2A and 2B) is
especially notable for \ion{Ca}{2} absorption and we highlight this in the
left panel of Figure \ref{stisccd_pt1}.  Slit 2B shows deep blueshifted
absorption of \ion{Ca}{2} 3934, 3968 \AA \ extending to $3790\unit{\AA}$
($-11,000$ km s$^{-1}$), whereas for Slit 2A  \ion{Ca}{2} absorption extends
only to $3835\unit{\AA}$ ($-7700$ km s$^{-1}$).  This suggests that the
remnant, although roughly spherical, is not isotropic in terms of absorption
line strengths, a fact also evident in the 2D spectra shown in
Figure~\ref{2D_All6Visits}.

Examining the 2D spectra for Slits 2A and 2B in detail gives a more nuanced
picture.  In the case of Slit 2B, the most extended \ion{Ca}{2} absorption
appears in the central region of the remnant, whereas in the spectrum of Slit
2A, the blueshifted side of \ion{Ca}{2} is most extended near the top of the
remnant while the red side appears to be more extended on the bottom (see
Fig.~\ref{2D_All6Visits}).  Determining the outermost
extent of the remnant's \ion{Ca}{2} absorption is made difficult due to the
overlap with the \ion{Ca}{1} absorption on its red side and  \ion{Fe}{1} 3720
\AA \ absorption on its blue side.

\subsection{A Clumpy Shell of \ion{Ca}{2} }

The presence of \ion{Ca}{2} clumps is shown in Figure~\ref{Ca_clumps}
where we present enlargements of the six spectra presented in
Figure~\ref{2D_All6Visits} stretched to reveal \ion{Ca}{2} absorption
variations.  Enhanced absorption at both blueshifted and redshifted velocities appears
present at all slit positions.  The strongest \ion{Ca}{2} absorptions are
seen to lie within an expansion velocity of $\pm$6000 km s$^{-1}$ for most
spectra, with few clumps centered zero velocity (the vertical line shown
in Fig.~\ref{Ca_clumps}).

A clumpy shell of \ion{Ca}{2} absorption out to roughly 6000 km s$^{-1}$ with
less absorption at low velocities is in qualitative agreement with high-resolution 
{\sl HST} images of the remnant.  Narrow-passband \ion{Ca}{2} images
of SNR~1885 show that the \ion{Ca}{2} absorption is strongest in a broad shell
with expansion velocities of $\simeq2000 - 6000 $ km s$^{-1}$
\citep{Fesen2007,Fesen2015}. Moreover, the reported $\simeq0\farcs05$ size of
the \ion{Ca}{2} clumps on these narrow \ion{Ca}{2} passband images
corresponding to $\simeq1500$ km s$^{-1}$ are consistent with the velocity
dispersion of the \ion{Ca}{2} clumps seen in the STIS/CCD spectra.  Below in
\S4, we discuss a model of SNR~1885's \ion{Ca}{2} absorption. 

\subsection{\ion{Ca}{1}}

Weak absorption longward of \ion{Ca}{2} $3934,
3968\unit{\AA}$ absorption was detected for all slits which we
attribute to \ion{Ca}{1} $4227\unit{\AA}$ in accord with \citet{Fesen1999}.
The central Slit 2B shows the broadest \ion{Ca}{1} absorption spanning from
$4120\unit{\AA}$ to $4420\unit{\AA}$ ($-7600$ to $+13,700$ km s$^{-1})$ while
the full range of blueshifted and redshifted \ion{Ca}{1} absorption velocities in all
six slits spans $-7900$ to $-6500$ and $+6600$ to $+13{,}700$ km s$^{-1}$ (see
Figs.~\ref{1DSpectra_all_positions} \& \ref{stisccd_pt1} and Table 2).  
It is important to note again that the blueshifted velocities may be
underestimates of the true radial velocity because of the overlap between the
\ion{Ca}{1} and \ion{Ca}{2} absorptions.

The remnant's \ion{Ca}{1} absorption is asymmetric, with stronger absorption on
the blue side of the line center compared to the red (Figs.
\ref{2D_All6Visits} \& \ref{2D_blue}) and 1D
(Figs.~\ref{1DSpectra_all_positions} \& \ref{stisccd_pt1}).  The weakest
\ion{Ca}{1} absorption appears in the southern (Slit 3B) and western (Slit 1A)
portions of the remnant, consistent with \ion{Ca}{1} images of the remnant
\citep{Fesen2007}. 

Except for Slit 3B, the line center of \ion{Ca}{1} feature appeared
redshifted by $\sim 1500$ km s$^{-1}$.  Based on a 1996 FOS spectrum, 
\citet{Fesen1999} also reported a redshift of $\sim
1100\unit{km}\unit{s}^{-1}$ for \ion{Ca}{1} absorption.
\citet{Fesen2007} argued that the redshifted \ion{Ca}{1}
absorption could not be accounted for by an anisotropic photoionization model
because there would be more ionization on the side closer to the bulge, resulting 
in a blueshifted \ion{Ca}{1} line profile.
 
\subsection{ \ion{Fe}{1}}
\label{Fe-sec}

A 1996 FOS spectrum of SNR~1885 showed \ion{Fe}{1} $3720\unit{\AA}$ absorption
on the blue wing  of the \ion{Ca}{2} absorption blend \citep{Fesen1999}.
Due to increased noise shortward of $4000\unit{\AA}$ caused by the
lower background bulge flux and the CCD's lower efficiency at blue wavelengths,
the presence of \ion{Fe}{1} is not immediately obvious from the 2D STIS/CCD
spectra shown in Figure \ref{2D_All6Visits}.
 
However, despite the increased noise in the spectra below 4000 \AA, \ion{Fe}{1}
absorption appears present in the 2D STIS spectra of 2A and 3A
(Fig.~\ref{2D_All6Visits}).  In fact, discrete blueshifted and redshifted \ion{Fe}{1}
3720 \AA \ absorption features can be seen in several of the 1D spectral plots
shown in Figure \ref{1DSpectra_all_positions} and for the two center Slits 2A
and 2B (Fig.~\ref{stisccd_pt1}, left panel).  In Table 2, we list the estimated
wavelengths and velocities of the \ion{Fe}{1} absorption features 
where present.

Despite the increased noise level below 4000 \AA, the appearance of matching
blueshifted and redshifted absorption features in the spectra for Slits 2A and
2B gives added weight to the reality of distinct \ion{Fe}{1} 3720 \AA \
absorption features.  For example, in the case of Slit 2A, blueshifted
\ion{Fe}{1} absorption appears present between $3600\unit{\AA}$ and
$3690\unit{\AA}$, and redshifted absorption between $3700\unit{\AA}$ and
$3830\unit{\AA}$.  For Slit 2B, the blueshifted \ion{Fe}{1} absorption also
appears around $3595\unit{\AA}$ and $3680\unit{\AA}$, and there is additional
absorption at $3700\unit{\AA}$ extending redward to at least $3790\unit{\AA}$
where it overlaps with the blue wing of \ion{Ca}{2} absorption (see Table 2).

\begin{figure*}[t]
    \begin{center}
    \leavevmode
   \includegraphics[scale=.7]{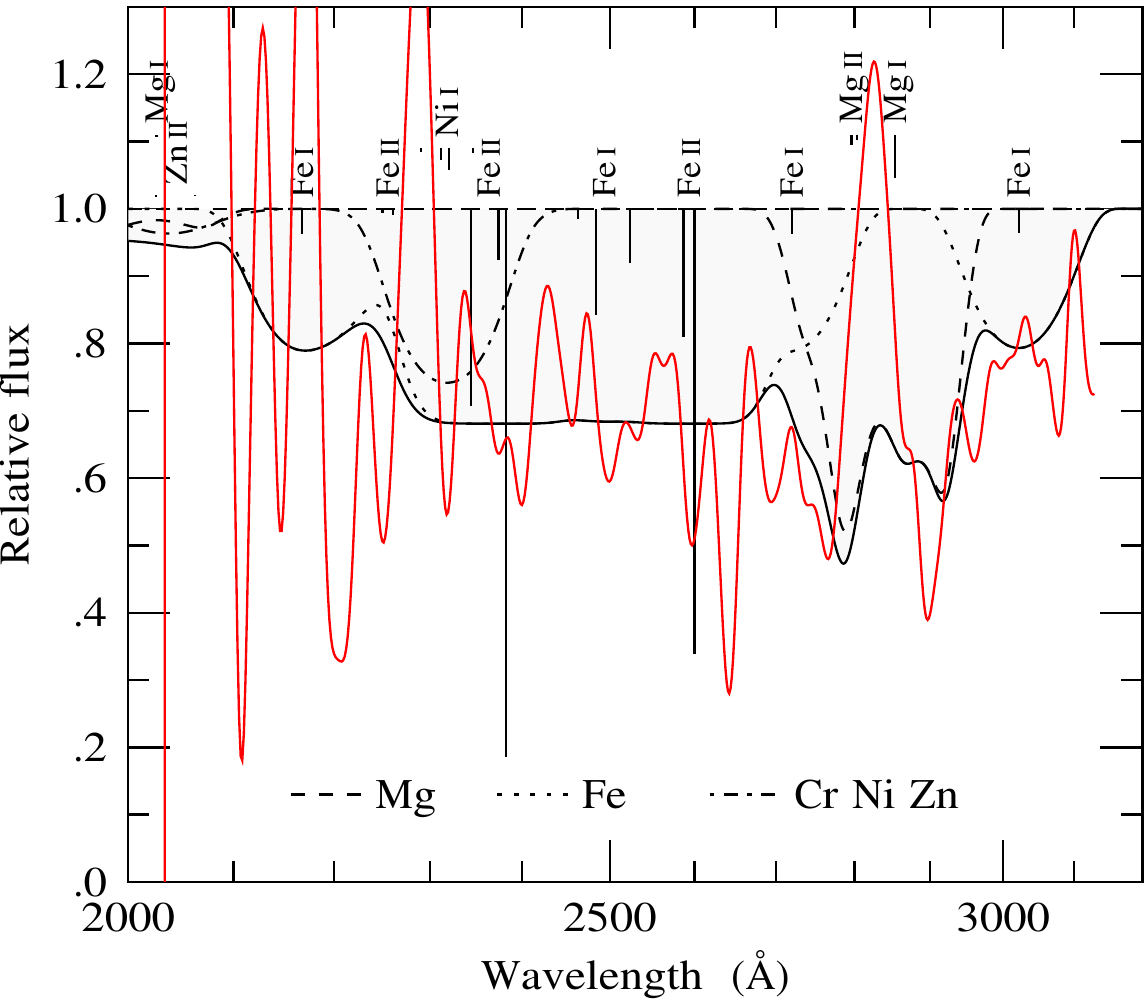}
    \hspace{2em}
   \includegraphics[scale=.7]{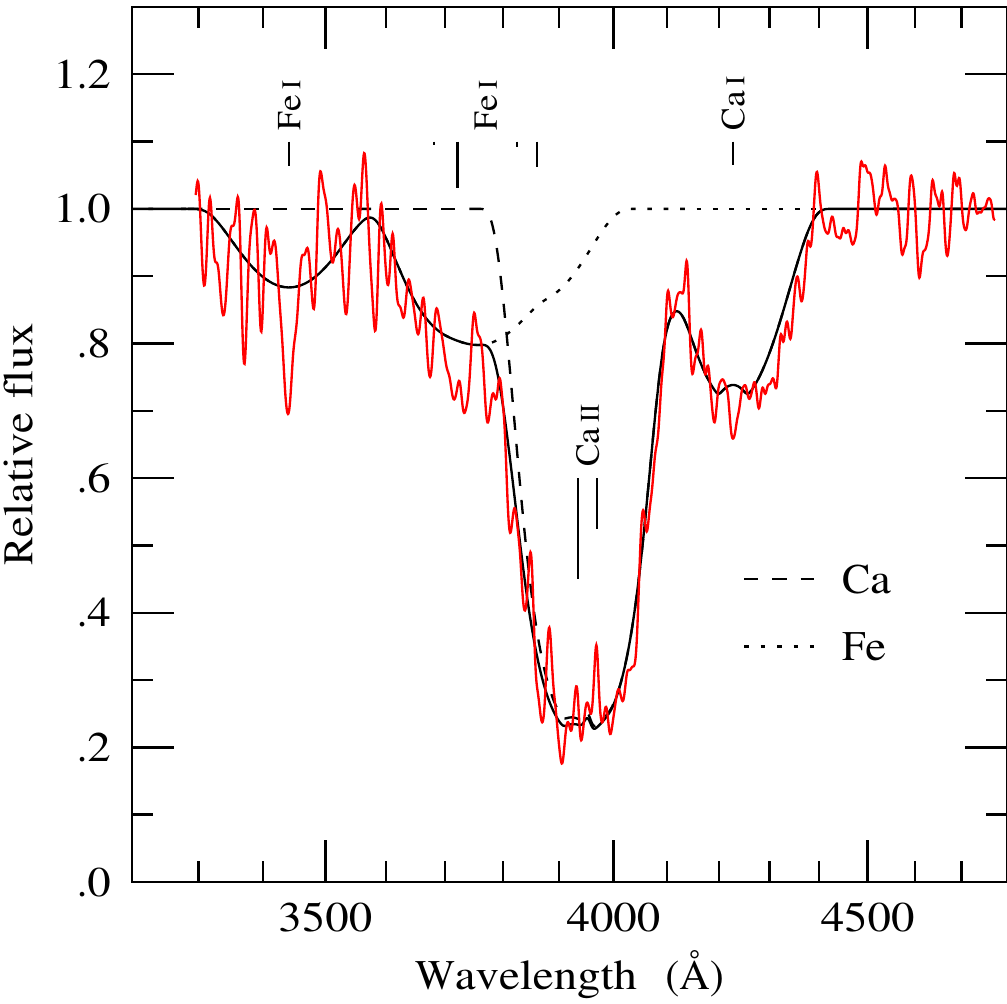}
    \end{center}
\caption{
\label{stismama} The UV STIS/MAMA spectrum for SNR~1885 (left panel, red line)
compared to a model fit (solid black line) 
used to match the optical FOS spectrum (right panel, red line) of \cite{Fesen1999}.
The W-shaped absorption feature centered near $2825 \unit{\AA}$
is modeled as a shell of \ion{Mg}{1} $2852 \unit{\AA}$
and \ion{Mg}{2} $2800 \unit{\AA}$ absorption.
Vertical lines,
with lengths proportional to optical depths
(oscillator strength times wavelength times elemental and ion abundance)
mark wavelengths of the strongest lines.
}
    \end{figure*}

The detection of separate blueshifted and redshifted \ion{Fe}{1} absorption
features suggests the presence of an \ion{Fe}{1} shell, an 
arrangement quite different from that suggested from inspection of the 1996 FOS
data,  which indicated a fairly symmetric absorption profile (right panel of
Fig.~\ref{stismama}).  Such a significant difference in the \ion{Fe}{1}
absorption profile between spectra taken nearly two decades apart may indicate
progressive ionization of \ion{Fe}{1} into \ion{Fe}{2} much like that predicted
and discussed by \citet{Fesen1999}.  Other potential \ion{Fe}{1} absorption
features lie further into the UV, and we discuss those in \S4.


\subsection{Mg I and Mg II}

Despite a low S/N, the UV spectrum of SNR~1885 shows depressed flux near the
wavelengths of \ion{Mg}{1} $2852 \unit{\AA}$ and \ion{Mg}{2} 2796, 2803 \AA.
The circles in the lower panel of Figure~\ref{stismama2D_BnW}  mark absorption
rings centered on these wavelengths. The size and thickness of these absorption
features, if attributed to a shell of Mg-rich ejecta, suggest material
expanding at velocities $\simeq$ 7000 to 10,000 $\unit{km} \unit{s}^{-1}$. We
note that if the observed absorption is associated only with \ion{Mg}{2} lines,
then they are blueshifted by $-700 \unit{km} \unit{s}^{-1}$ relative to the
Earth rest frame. 

A detection of a Mg-rich shell adds more to the study of SNR~1885 than
just the addition of Mg to the already detected Fe and Ca.  
The remnant's Mg absorption shell exhibits a 
higher minimum velocity compared to the broad shell of Ca-rich
ejecta and the sparse finger-like structure seen for the remnant's \ion{Fe}{2}
material \citep{Fesen2015}.

    \begin{figure*}[t]
    \begin{center}
    \leavevmode
    \includegraphics[scale=.7]{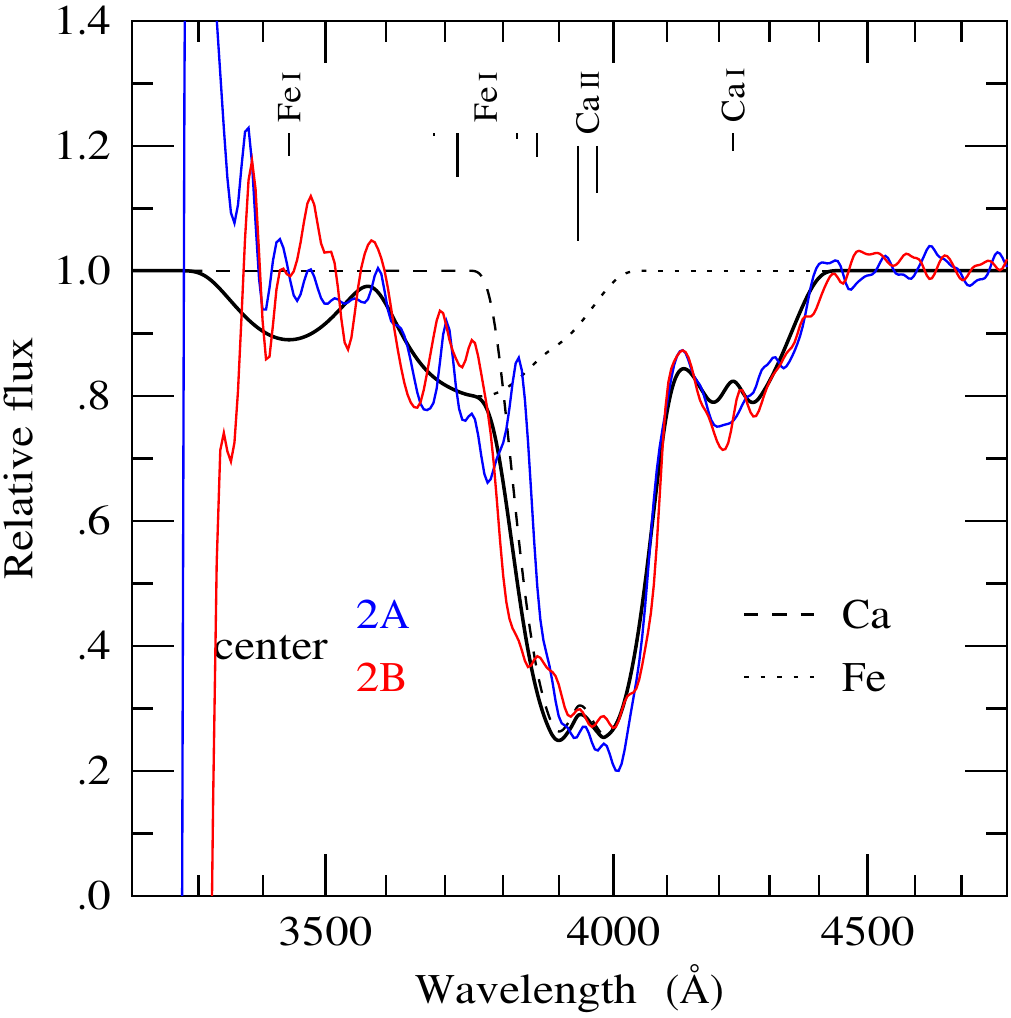}
    \hspace{2em}
    \includegraphics[scale=.7]{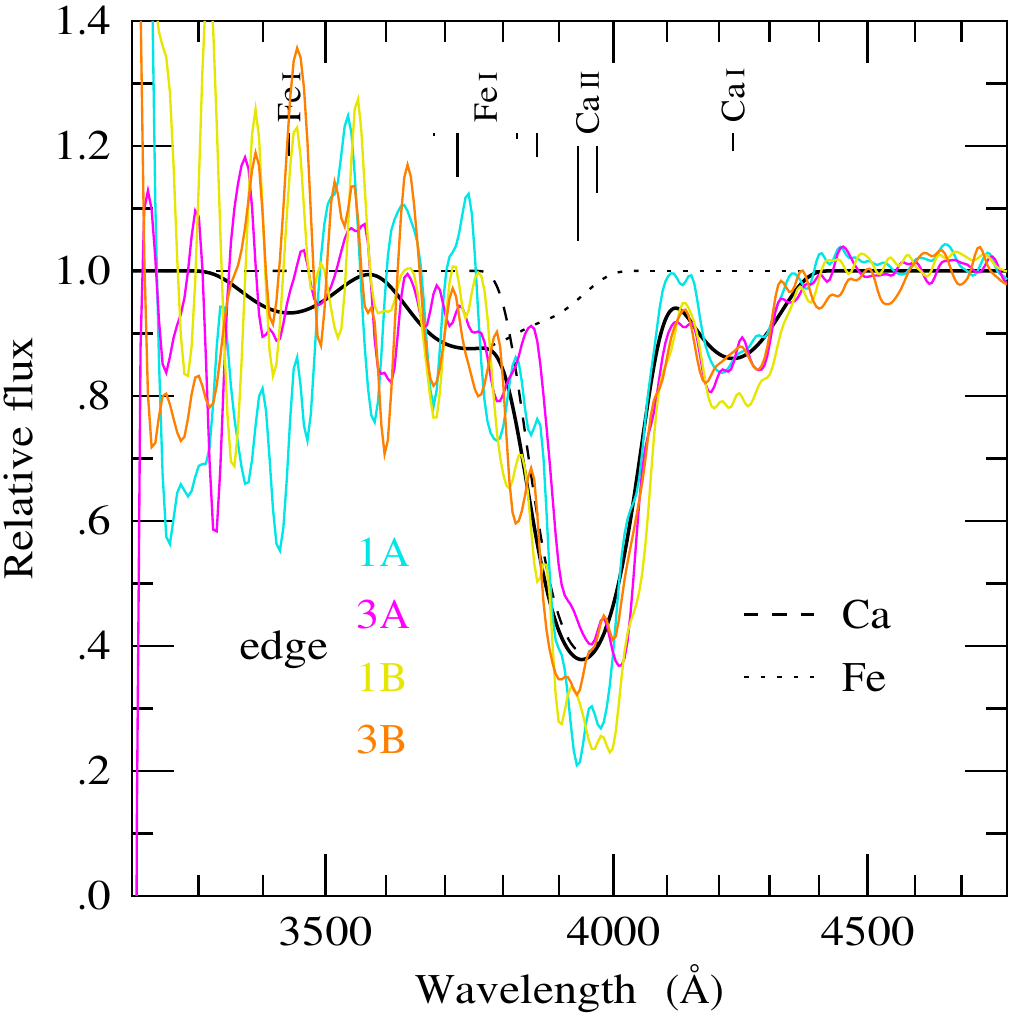}
    \caption{
\label{stisccd_pt2}
Background-corrected STIS/CCD spectra
for the central 3 pixels ($0\farcs3$) of the slit
for each of the six visits.
The panels show the observed spectra along with our model fit
(black lines).
    }
    \end{center}
    \end{figure*}



\section{Discussion}


Figure~\ref{stismama} shows the observed 1D UV spectrum of SNR~1885 compared to the 
model spectrum (left panel) used to fit the 1996 optical FOS spectrum (right
panel) discussed by \cite{Fesen1999}.  The model shown here for the UV spectrum
differs from that in \cite{Fesen1999} in a number of details. 

As before, we assume spherical symmetry given the approximate spherical
symmetry seen both in the images \citep{Fesen2015} and in the 2D spectra (e.g., see
Figure~\ref{2D_blue}).  However, the remnant cannot be exactly spherically
symmetric; if it were, then the two spectra for the central slices in the
left panel of Figure~\ref{stisccd_pt1} would appear the same, as would the four
spectra for the edge slices.  Moreover, the remnant's spherical structure is
further complicated by ejecta clumping as noted above.

Before discussing the model results, we note that the observed level of absorption
over the $2200$ -- $3200$ \AA \ wavelength region is much weaker than anticipated by the
model described by \citet{Fesen1999}.  The predicted level of absorption in that
work was based on the expectation that \ion{Fe}{2} should be substantially more
abundant than \ion{Ca}{2} in this Type~Ia remnant, which was seen to be both
broad and deep in the FOS spectrum \citep{Fesen1999}.  Indeed,
the flattish bottom of the \ion{Ca}{2} absorption profile (right panel of
Fig.~\ref{stismama}) suggested that \ion{Ca}{2} absorption was saturated, with
the residual level of flux arising from starlight in the bulge of M31 to the
foreground of SNR~1885.

If Fe were distributed similarly to that of Ca, and if \ion{Fe}{2} were at least
as abundant as \ion{Ca}{2}, then the remnant should show similarly saturated
absorption in the resonance-line complexes of \ion{Fe}{2} at $2380 \unit{\AA}$
and 2600 \AA.  However, the observed UV spectrum of SNR~1885, shows absorption
levels less than half as deep as expected.

\masstable

\subsection{Calcium}

The flat-bottomed profile of the observed \ion{Ca}{2} H \& K absorption seen
both in the FOS and STIS spectra suggests that \ion{Ca}{2} is nearly saturated.
Based on a narrow-passband \ion{Ca}{2} image of the remnant, \citet{Fesen2015}
argued that \ion{Ca}{2} is densest in a lumpy shell over $2000$ -- $6000$
km s$^{-1}$. This is consistent with the clumpy \ion{Ca}{2} absorption 
seen in the STIS/CCD spectra (Fig.\  \ref{Ca_clumps}).
 
We therefore adopted a model similar to that described in \cite{Fesen1999},
in which the intrinsic density profile
$n(v)$
of all elements is taken to be the same
as that inferred from the line profile of \ion{Ca}{2},
\begin{equation}
\label{nv}
  n(v)
  \propto
  \left[ 1 - ( v / v_{\max} )^2 \right]^2
  , \quad 
  v < v_{\max} = 13{,}400 \unit{km} \unit{s}^{-1}
  \ ,
\end{equation}
but with a hole defined by a sharp minimum velocity
$v_{\min} = 2000 \unit{km} \unit{s}^{-1}$. 
Since the data are not high enough quality 
to say otherwise, we also assume that \ion{Ca}{1} has the same density
distribution as \ion{Ca}{2}.  

This model assumes the Sobolev optical depth approximation (see Eq.\ 54 in
\citealt{Jeffery1993} where $\beta = $ v/c).  Use of this approximation is
valid since the change of the physical conditions over the coupling region
between the line and the photons is small, meaning  the scale of the change is
roughly the intrinsic line width compared to the expansion velocity.  The
relevant regions of SNR~1885 are still in free expansion with typical
velocities of 1000 km s$^{-1}$ or more that are larger than the intrinsic line
widths by many orders of magnitude.

If there is no hole in the Ca distribution, then the flat-bottomed
\ion{Ca}{2} profile points to \ion{Ca}{2} being saturated.  The non-zero flux
at line bottom would then be attributed to foreground starlight from the bulge
of M31, with the foreground starlight fraction being $0.21$.
However, the case with no foreground starlight is only 3-$\sigma$ from
the best fit.

With a central hole, \ion{Ca}{2} can be less saturated, meaning less
\ion{Ca}{2}, and the foreground starlight fraction can be lower.  A lesser
amount of \ion{Ca}{2} also translates into relatively less absorption in the
spectra of edge slices compared to central slices.  Less absorption in edge
versus central slices is in fact what is observed with STIS/CCD.  

Figure~\ref{stisccd_pt2} shows the observed optical spectrum of SNR~1885
compared to the model spectrum used to fit the 1996 optical FOS spectrum for
both the two center slits (left panel) and the four edge slits (right panel).
The model shown in both Figures~\ref{stismama} and~\ref{stisccd_pt2} adopts a
foreground starlight fraction of $0.16$, the 1-$\sigma$ low end of the value
measured from the FOS.

Allowing for a hole in the calcium distribution also leads to slight changes in
other parameters from those cited in \cite{Fesen1999}.  One effect concerns the
degree of ionization of calcium. The original fit to the FOS data suggested
\ion{Ca}{2}/\ion{Ca}{1} $= 16$, however our new model fit suggests a lower
value of 13.  However,  the degree of ionization measured from the STIS/CCD
spectra is 16.  Although the absolute level of ionization is model dependent,
the conclusion that the degree of ionization is higher in the STIS/CCD spectrum
than in FOS is robust, since it depends only on the fact that \ion{Ca}{1}
absorption is somewhat weaker in STIS/CCD than in FOS.

This degree of ionization of calcium is consistent with theoretical
expectations.  As discussed by \cite{Fesen1999}, the ionization timescale of
\ion{Ca}{1} exposed to ultraviolet light from the bulge of M31 is only $8
\unit{yr}$.  Such a short ionization time would imply that a value of
\ion{Ca}{2}/\ion{Ca}{1} $= 13$ at the epoch of the FOS observation would
translate to a predicted \ion{Ca}{2}/\ion{Ca}{1} $= 19$ at the epoch of
the STIS/CCD observations.

Table~\ref{masstable} lists the parameters of the model of the present paper, along
with the parameters of the model in \cite{Fesen1999}.  We note that in order
for sufficient \ion{Ca}{1} survive to the present time, it must be at least
partially self-shielded by its own continuum photoionization cross-section,
thus slowing the rate of photoionization. 

\subsection{Iron}

The model fit for the STIS/MAMA spectrum shown in Figure~\ref{stismama} takes into
account the expected $t^{-2}$ reduction in time $t$ of the column density of a
freely-expanding supernova.  That is, column densities in the
STIS/MAMA and STIS/CCD spectra, taken when the remnant was 128.3 years old, have
been reduced by $( 128.3 \unit{yr} / 111.4 \unit{yr} )^{-2} = 0.754$ compared
to the model for the FOS spectrum at an age of 111.4 years.

A puzzle concerns the observed level of Fe absorption relative to
\ion{Ca}{2} H \& K. To address this, it is useful to first address the strength
of \ion{Fe}{1} absorption.

The strength and distribution of \ion{Fe}{1} absorption in the FOS
spectrum, shown in the right panel of Figure~\ref{stismama}, is different from
that seen in the STIS/CCD spectra, where it appears to be in the form of a shell.
In addition, a few other \ion{Fe}{1} absorption lines lie within 
the wavelength range of STIS/MAMA (see Fig.\ \ref{stismama}, left panel).

The strongest resonance line of \ion{Fe}{1} over the STIS/MAMA range is the
four line complex around 2500 \AA, consisting of 2463, 2483, 2501, 2522 \AA,  with
a combined oscillator strength of $0.939$ \citep{Morton91}.  By comparison, the
\ion{Fe}{1} $3720 \unit{\AA}$ blend (3679 \& 3720 \AA) has a combined
oscillator strength of just $0.044$.  Since optical depths in the Sobolev
approximation are proportional to oscillator strength times wavelength, the
optical depth in \ion{Fe}{1} $2500 \unit{\AA}$ should be more than 10 times
that of \ion{Fe}{1} $3720 \unit{\AA}$.

Given that the optical depth of \ion{Fe}{1} $3720 \unit{\AA}$ observed with FOS
was $\approx 0.25$ at line center, the expectation is that UV \ion{Fe}{1}
absorption around 2500 \AA \ would be saturated, reaching a depth of $\approx
0.8$, similar to that seen for \ion{Ca}{2} (i.e., the observed intensity is
$0.2$ of continuum coming from foreground M31 bulge starlight). 

However, the observed level of \ion{Fe}{1} $2500 \unit{\AA}$ absorption is just
$\approx 0.3$, in conflict with expectations by a factor of 2 -- 3.  The problem
is made worse since there are strong resonance lines of  \ion{Fe}{2} in the UV
STIS/MAMA spectrum that lie at 2343, 2374, 2382, 2586, 2599 \AA, with a
combined oscillator strength of $0.727$. 

The ionization fraction of iron adopted in the model of Figure~\ref{stismama}
is \ion{Fe}{2}/\ion{Fe}{1} $= 10$, the same as that adopted by
\cite{Fesen1999}.  This was chosen to be consistent with the ionization
fraction of calcium, \ion{Ca}{2}/\ion{Ca}{1} $= 16$, measured from the STIS/CCD
spectra. The model predicts that the \ion{Fe}{2} absorption lines in the
STIS/MAMA spectrum would be completely saturated. However, as can be seen in
Figure~\ref{stismama}, the observed level of absorption is
far from being saturated.

One might question the validity of the STIS UV data. However, the fact that
there is good consistency between the individual spectra indicates there is no
significant problem with the observations.  Consequently, we 
conclude that the weakness of \ion{Fe}{1} and \ion{Fe}{2} absorption
observed with STIS/MAMA requires a covering fraction less than 1, 
that we assumed for \ion{Ca}{2}.

The idea that \ion{Fe}{2} has a covering fraction less than 1 would be
consistent with the \ion{Fe}{2} F225W image reported by
\citet[Fig.~7]{Fesen2015}. This image shows \ion{Fe}{2} concentrated in 4
plumes extending out from the center, with implied velocities $\sim 10{,}000
\unit{km} \unit{s}^{-1}$ (see Fig.~\ref{Elements}). This is in sharp
contrast to the more spherical appearance of \ion{Ca}{2}.  

The remnant's \ion{Fe}{1} distribution likewise appears nonspherical.  
Narrow-passband \ion{Fe}{1} images \cite[Fig.~5]{Fesen2015} sensitive to $\leq$~5000 km
s$^{-1}$ expansion velocities show an off-center (hence nonzero) blob
of absorption. This arrangement is consistent with the STIS/CCD spectra that 
show the remnant's \ion{Fe}{1} absorption largely confined to nonzero
velocities, ranging from $2000$ to 8000 km s$^{-1}$.

With the foreground starlight fraction fixed at $0.16$, the UV STIS/MAMA
spectrum, coupled with the level of \ion{Fe}{1} 3720 \AA \ observed by FOS and
STIS/CCD, requires that iron, in the forms of \ion{Fe}{1} and \ion{Fe}{2}, has
a covering factor of $0.38 \pm 0.07$. 
The model shown in Figure~\ref{stismama} assumes this covering fraction
for Fe and other iron-group elements, including Cr, Ni, and Zn.

This means 62\% of light from behind the remnant is leaking through iron-group
elements unimpeded.  While the exact iron covering factor is correlated with the
foreground starlight fraction, the STIS/MAMA spectrum constrains the net level
of Fe absorption to be $(1 - 0.16) \times 0.38 \pm 0.07 = 0.32 \pm 0.06$.

To compensate for the decreased covering fraction for iron while still
reproducing the level of \ion{Fe}{1} absorption seen in the FOS spectrum (right
panel of Figure~\ref{stismama}), the abundances of iron-group elements have
been increased by a factor of 3 compared to the model of \cite{Fesen1999}.  The
abundance of Mn, which has a strong resonance line complex \ion{Mn}{1} $2800
\unit{\AA}$ for which there is no evidence in the observed STIS/MAMA spectrum,
has been set to zero.

The fit of our new model to the \ion{Fe}{1} FOS profile is not quite as good as
that found by \cite{Fesen1999} who assumed the same covering fraction for Fe
as Ca, but it is acceptable.  The mass of \ion{Fe}{1} in the model is $0.05 \pm
0.01 \Msun$ (1-$\sigma$ statistical uncertainty), a factor 3 increase over that
in \cite{Fesen1999}.

At an \ion{Fe}{2}/\ion{Fe}{1} ionization fraction of 10, the mass of \ion{Fe}{2}
is $\sim 0.5 \Msun$.  However, this estimate is highly uncertain, because the
model predicts that the \ion{Fe}{2} absorption lines in the STIS/MAMA spectrum
would be completely saturated, preventing a direct measurement of
\ion{Fe}{2}/\ion{Fe}{1}.

Finally, \citet{Fesen1999} estimated that the continuum optical depth of
\ion{Fe}{1} through the center of the remnant is $\sim 0.1$ at wavelengths
shorter than the $1569 \unit{\AA}$ ionization threshold of \ion{Fe}{1}.  The
increase of a factor of 3 in Fe in the present analysis implies a continuum
optical depth of $\sim 0.3$.  This modest optical depth suggests that the
remnant should currently be undergoing a transition from being optically thick
to being optically thin in \ion{Fe}{1}.  A corollary of this conclusion is that
\ion{Fe}{2} should also be present in regions where there is \ion{Fe}{1}.

\subsection{A Magnesium-Rich Shell}
\label{Mg-sec}

The new and most striking feature of the STIS/MAMA spectrum, besides the unexpected
weakness of Fe absorption, is the apparent shell of absorption near $2800
\unit{\AA}$ evident in the 2D spectrum in Figure~\ref{stismama2D_BnW}. This
appears as a W-shaped absorption feature in the 1D spectrum in
Figure~\ref{stismama}.  While the spectrum is quite noisy, this absorption is
present in all visits.

The model spectrum plotted in Figure~\ref{stismama} attempts to fit the feature with a
spherically symmetric shell of \ion{Mg}{1} $2852 \unit{\AA}$ and \ion{Mg}{2}
$2800 \unit{\AA}$ absorption.  The Mg shell is assumed to have the same density
profile $n(v)$, equation~(\ref{nv}), as Ca, but with a sharp minimum velocity
taken to be $v_{\min} = 7000
\unit{km} \unit{s}^{-1}$. 

    \begin{figure}
    \begin{center}
    \leavevmode
    \includegraphics[scale=.6]{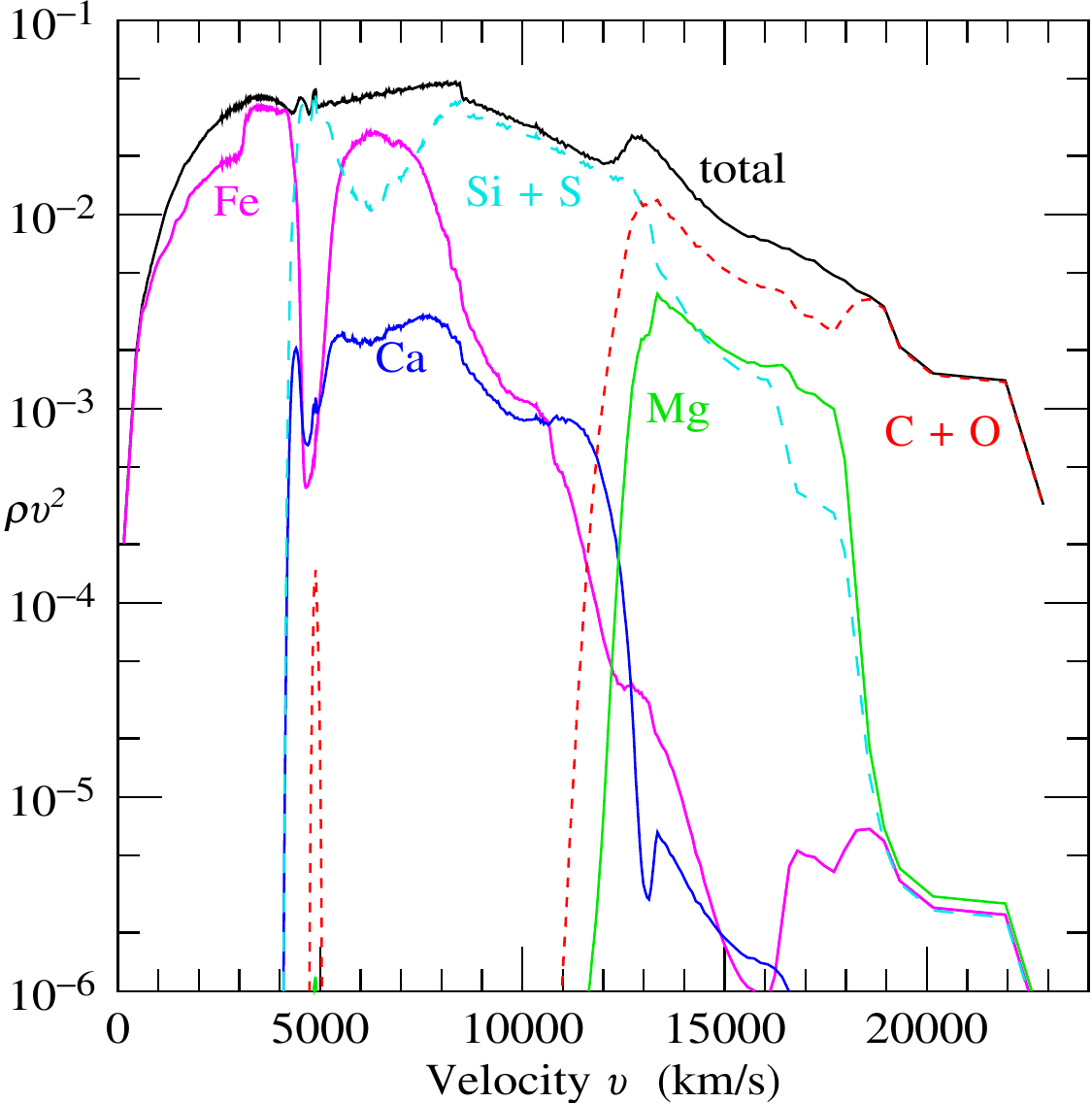}
    \caption{
\label{snmodel}
Density $\rho$ times velocity $v$ squared
as a function of expansion velocity $v$
from a spherical delayed-detonation model 5p02822.16
\citep{Hoeflich02}.
    }
    \end{center}
    \end{figure}

The ionization time of \ion{Mg}{1} exposed to
photoionizing light from the bulge of M31 is $\approx 44 \unit{yr}$, as
compared to the $\approx 8 \unit{yr}$ photoionization time of \ion{Ca}{1}
\citep{Hamilton00}. Consequently, \ion{Mg}{1} is more robust than \ion{Ca}{1} and
hence less ionized. 

If the time of photoionization of \ion{Mg}{1} is the same as that of \ion{Ca}{1}, then
the ionization fraction of \ion{Mg}{2}/\ion{Mg}{1} is $\approx 0.7$.  This
means that there are almost equal abundances of neutral and singly-ionized
magnesium.  The center of the observed Mg absorption shell appears to lie about
halfway between \ion{Mg}{1} $2852 \unit{\AA}$ and \ion{Mg}{2} $2800
\unit{\AA}$.  Having no theoretical or observational reason to prefer
\ion{Mg}{1} or \ion{Mg}{2}, we allow both in the model, setting
\ion{Mg}{2}/\ion{Mg}{1} $= 0.7$ consistent with the degree of ionization of Ca.

However, the model's shell of Mg fails to fit the observed hole at the shell's center,
showing scarcely a central dip in absorption, let alone a gaping hole.
The flaw in the model is probably due the assumption of spherical symmetry. In
fact, the 2D spectrum in Figure~\ref{stismama2D_BnW} indicates that the shell
is most prominent on the near and far sides of SNR~1885, and weaker at central
velocities.  Thus, we conclude that there is evidence for a Mg shell but that is 
may well be incomplete and lumpy.

    \begin{figure*}[t]
    \begin{center}
    \leavevmode
    \includegraphics[scale=0.9]{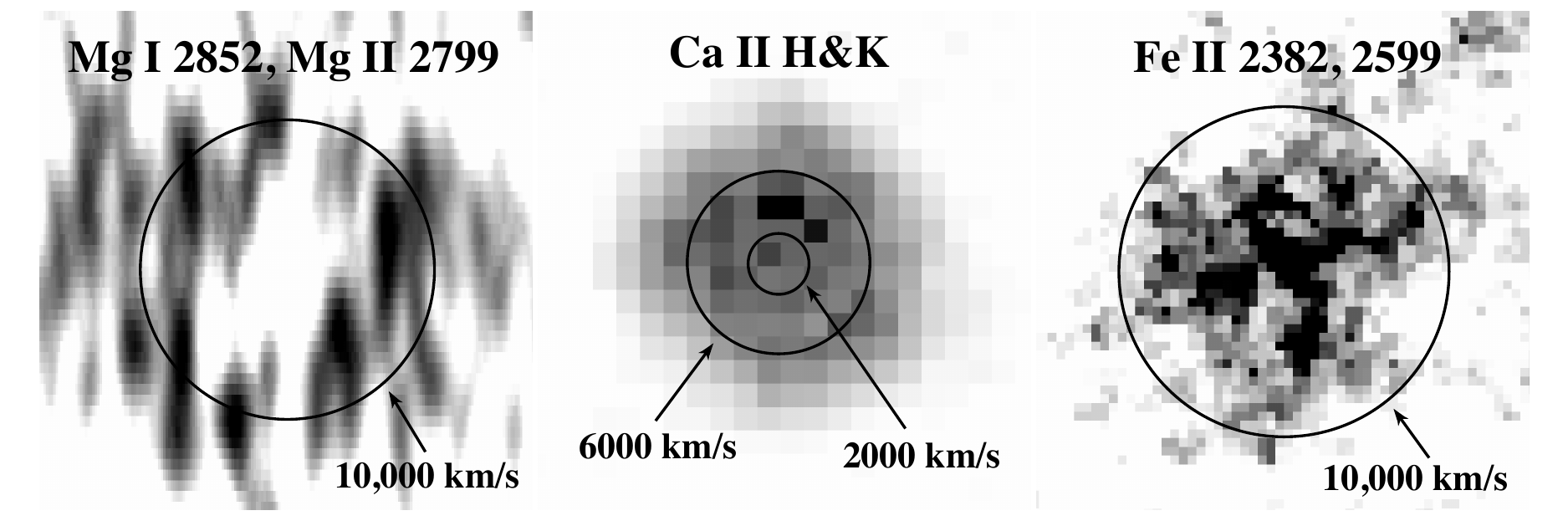}
    \caption{
\label{Elements}
Observed absorption distributions of \ion{Mg}{1} + \ion{Mg}{2}, \ion{Ca}{2}, and \ion{Fe}{2} rich 
ejecta based  
UV spectra for Mg and {\sl HST} images of Ca and Fe
from \citet{Fesen2015}. Scale: 10$^{4}$ km s$^{-1}$ $\simeq 0\farcs3$. 
    }
    \end{center}
    \end{figure*}

\section{SN~Ia Models and the Remnant of SN~1885}
\label{SNIa-sec}

\subsection{Delayed-Detonation Models}

The details of SNe~Ia explosions are complicated and the subject of much
debate.  Numerical calculations indicate that the thermonuclear runaway of a
near Chandrasekhar-mass white dwarf is preceded by a century-long, highly
turbulent, convective, carbon-burning, ``smoldering'' phase
\citep{Sugimoto80,Nomoto82,HN00}.  Ignition, which occurs at around $1.5 \times
10^{9} \unit{K}$, is highly sensitive to temperature and density, and may occur
off-center and proceed along paths of least resistance determined by the
fluctuating conditions around the ignition point
\citep{Timmes1992,Garcia1995,Niemeyer1996,hs02,Livne05}.

Burned regions expand and are Rayleigh-Taylor unstable against overturn with
denser overlying matter. Three-dimensional calculations show
that the subsonic deflagration proceeds mainly along Rayleigh-Taylor unstable
radial fingers, leaving behind connected regions of unburned matter \citep{gamezo2005}. 

The initial deflagration heats the white dwarf, causing it to expand
quasi-hydrostatically over a period of about a second.  At that point, the
deflagration somehow turns into a supersonic detonation that propagates through
the white dwarf, consuming unburned material.  Near the center of the white
dwarf, densities and temperatures are high enough that the burning reaches
nuclear statistical equilibrium.  But farther out, the expansion of the white
dwarf prevents nuclear burning from continuing to completion, so that
intermediate-mass elements are produced, with less-massive elements being
produced farther out.

During both the initial deflagration phase and the subsequent detonation phase,
Rayleigh-Taylor instabilities lead to radial mixing.  Mixing proceeds at a rate
that depends on the sound speed and continues until the white dwarf reaches
homologous expansion, which occurs about 10 -- 20 seconds after the explosion.
Three-dimensional hydrodynamic calculations suggest that the Rayleigh-Taylor
instabilities are strong enough to mix the remnant's entire central region
\citep{gamezo03,r02,roepke12}.

However, full radial mixing is inconsistent with the layered structure of
intermediate-mass elements that is seen both in SN~Ia spectra and in the
observations of SNR~1885.  Somehow mixing must be suppressed.  A
possible mechanism to suppress Rayleigh-Taylor mixing is high magnetic fields
\citep{Hoeflich2013,Penney2014,Diamond15}.

A simple, albeit artificial, way to suppress mixing is to impose spherical
symmetry.  Figure~\ref{snmodel} shows an example calculation from a suite of
spherically symmetric computations of delayed-detonation explosions by
\cite{Hoeflich02}. The model, 5p02822.16, is subluminous compared to normal
SN~Ia.  The models have two free parameters, the central density at which the
explosion begins and the density at which the deflagration transitions into a
detonation.  The central density is determined essentially by the accretion
rate, and it determines the amount of nickel-iron synthesized in the explosion.
A higher accretion rate leads to a higher central density and greater
production of nickel-iron \citep{HWT98}.

\subsection{The Remnant of SN~1885}

Figure~\ref{Elements} shows a comparison of the elemental distribution of
Mg-rich ejecta based on the STIS/MAMA spectrum and \ion{Ca}{2} and \ion{Fe}{2}
based on direct images \citep{Fesen2015}.  The full extent of the remnant's
Ca-rich ejecta as seen in {\sl HST} images reaches a radius of $0\farcs80$
(13,400 km s$^{-1}$).  The striking appearance of \ion{Fe}{2} plumes in
SNR~1885 in Figure~\ref{Elements}, would seem to support the presence of
Rayleigh-Taylor fingers.

The angular size of the remnant's Ca absorption is consistent with the observed
velocity of Ca absorption. This implies that SNR~1885's ejecta are nearly
in free expansion.  Consequently, the distribution of its elements is
essentially the same as that established a few seconds after the supernova
explosion. Thus SNR~1885's elemental structure is a powerful tool for
investigating the internal structure of a SN~Ia.  

The key salient features of the observed density distribution are as follows.

1) The UV STIS/MAMA spectrum of SNR~1885, \S\ref{Fe-sec}, requires that iron,
both in the form of \ion{Fe}{1} and \ion{Fe}{2}, has a covering fraction of
only $\approx 0.38$.  This suggests that iron is mainly confined to dense,
optically thick regions that cover only a fraction of the remnant. This is
consistent with iron being concentrated in a few plumes, as seen in WFC3/F225W
images (Fig.\ \ref{Elements}; \citealt{Fesen2015}).  The STIS/CCD spectra
suggests that \ion{Fe}{1} has a shell-like structure extending roughly from 2000 to
8000 km s$^{-1}$. 

2) Although \ion{Ca}{2} absorption is present from zero velocity to $\approx
13{,}000 \unit{km} \unit{s}^{-1}$, \ion{Ca}{2} appears to be densest in a broad
shell expanding at $2000$ -- $6000 \unit{km} \unit{s}^{-1}$.  This shell of
\ion{Ca}{2} is evident both in narrowband imaging of \ion{Ca}{2} reported by
\cite{Fesen2015}, reproduced in Figure~\ref{Elements}, and in the STIS/CCD
spectral data shown in Figure~\ref{Ca_clumps}. Some SN~Ia models show a
destruction of Ca in the progenitor's densest inner region
\citep{Hoeflich02,HWT98}, which could lead to a Ca-rich shell like that
observed.

\ion{Ca}{1} absorption extends over the same range of velocities
as \ion{Ca}{2} (see Table 2).  The total observed mass of calcium, mostly in
the form of \ion{Ca}{2}, is $0.004 \Msun$.  The mass of calcium could possibly
be larger than this if, like iron, some calcium is hiding in optically thick
lumps.

3) The UV STIS/MAMA spectrum of SNR~1885 indicates that magnesium, either in
the form of \ion{Mg}{1}, \ion{Mg}{2} or both, is concentrated in a shell that
starts at about $7000 \unit{km} \unit{s}^{-1}$ and extends out to at least
10,000 km s$^{-1}$.  The total observed mass of magnesium is roughly $0.03
\Msun$.

\subsection{Implications of SNR~1885 for SN~Ia Models}

Our spectral results, together with the recent imaging data
of \cite{Fesen2015}, are largely consistent with the implications of SN~Ia models
discussed by \cite{Fesen2007}.  What is new is the
observational evidence for a few extended plumes of iron \citep{Fesen2015}, and
the detection of a shell of magnesium at velocities $\gtrsim 7000$ km s$^{-1}$.

 \snmasstable

Figure~\ref{snmodel} shows the density $\rho$ of the subluminous SN~Ia 
model, 5p02822.16, as a function of velocity $v$ for the principal elements 
synthesized in the model explosion.  The density is shown multiplied by 
velocity squared, in part to compensate for the steep decline in density with 
velocity and in part because mass is proportional to the integral 
$\int \rho v^2 \dd v$ under the curve. A higher central density results 
in greater production of stable Ni.

The model in Figure~\ref{snmodel} shows Ca over $\approx 4000$ -- $12{,}500
\unit{km} \unit{s}^{-1}$, and Mg over $\approx 12{,}500$ -- $18{,}000 \unit{km}
\unit{s}^{-1}$.  Notice that the shell of Mg starts where the shell of Ca
stops.  

In contrast, the observations of SNR~1885 show Ca extending down to zero
velocity (albeit concentrated over $\approx 2000$ -- $6000 \unit{km}
\unit{s}^{-1}$), with Mg extending down to $\sim 7000 \unit{km} \unit{s}^{-1}$.
There appears to be a region $\approx 7000$ -- $13{,}000 \unit{km}
\unit{s}^{-1}$ in SNR~1885 where Ca and Mg mix and/or overlap.  This indicates
some radial mixing of these elements, but not as much as 3D 
hydrodynamic calculations would suggest.  

However, the model shows Mg at velocities up to $18{,}000 \unit{km} \unit{s}^{-1}$.
While there are possible hints of Mg in SNR~1885 at velocities exceeding
the maximum $\approx 13{,}000 \unit{km} \unit{s}^{-1}$ velocity of Ca,
there is no clear sign of Mg at velocities as high as in the model.

The model in Figure~\ref{snmodel} also shows substantial Si and S over $\approx
4000$ -- $13,000 \unit{km} \unit{s}^{-1}$, approximately the same velocity
range as Ca.  Si and S have strong UV absorption lines over $1200$ --  $1900
\unit{\AA}$ \citep{Fesen1999}, where the STIS/MAMA data are too noisy.  

Table~\ref{snmasstable} shows the masses of principal elements synthesized in
the subluminous model 5p02822.16 of Figure~\ref{snmodel}.  For comparison, the
Table also shows masses synthesized in 5p02822.25, a normal SN~Ia.  Whereas the
subluminous model yields $0.4 \unit{\Msun}$ of iron, a normal SN~Ia has  $0.7
\unit{\Msun}$ of iron.

The mass $0.03 \unit{\Msun}$ of Mg produced in the subluminous model is the
same as our Mg mass estimated for SNR~1885 (Table~\ref{masstable}), which is
higher than the $0.01 \unit{\Msun}$ for a normal SN~Ia model.  On the other
hand, a Ca mass of $0.04 \unit{\Msun}$ in subluminous or normal models is a
factor of 10 higher than the $0.004 \unit{\Msun}$ Ca mass we estimate in
SNR~1885.  This discrepancy seems to be a serious one, as there is no natural
way to dial down the mass of Ca synthesized in the models.

It is possible that SNR~1885 contains more Ca than our empirical model fit to
the observed \ion{Ca}{2} blend.  It could be more optically thick than we have
estimated.  Another possibility is that most of the remnant's Ca has
been ionized to \ion{Ca}{3} or higher.  

\ion{Ca}{2} is the only singly ionized species among abundant elements to have
an ionization potential, $11.87 \unit{eV}$, below the Lyman limit of $13.6
\unit{eV}$.  A calculation similar to that reported by \cite{Hamilton00} shows
that the photoionization timescale of \ion{Ca}{2} against photoionization by UV
light from the bulge of M31, as observed with IUE \citep{Burstein1988} and down
to the Lyman limit with HUT \citep{FD93}, is $\approx 200 \unit{yr}$.  This is
too long to produce substantial \ion{Ca}{3}.  While it is possible that reverse
shock-generated ultraviolet and X-ray emission could be photoionizing Ca-rich
ejecta to higher ionization stages \citep{HF88}, the presence of high-velocity
\ion{Ca}{1} absorption would seem to make this unlikely.

\section{Conclusions}

This paper has presented optical and UV imaging spectroscopy of the remnant of
SN~1885 taken during 2013 and 2014 with STIS on {\sl HST}.  These observations
clarify the distribution of the ejecta in this subluminous Type~Ia supernova
remnant. Key findings are as follows.

-- Optical spectra taken at several different slit positions in the remnant lend
supporting evidence for calcium to be present at all velocities from zero to
$\approx 13{,}400 \unit{km} \unit{s}^{-1}$ but densest in a clumpy shell with
expansion velocities mainly between 2000 and 6000 km s$^{-1}$.

-- Compared to a 1996 FOS spectrum, variations in the remnant's \ion{Fe}{1} and
\ion{Ca}{1} absorption strengths and line profiles are seen, in line with
predictions of ionization rates.

-- The remnant's overall UV spectrum shows absorption over $2200$ -- $3200
\unit{\AA}$ consistent with the expectation of absorption dominated by
\ion{Fe}{1} and \ion{Fe}{2} but at a level some 2 to 3 times weaker than
previously predicted \citep{Fesen1999}.  This suggests that the remnant's Fe-rich
ejecta is confined primarily to dense, optically thick regions with a covering
factor of just $\sim 0.4$, which is consistent with recent {\sl HST} images
showing \ion{Fe}{2} concentrated in 4 narrow plumes extending from the
remnant's center out to $\approx 10{,}000 \unit{km} \unit{s}^{-1}$
\citep{Fesen2015}.

-- UV spectra reveal the presence of a
shell of magnesium-rich ejecta expanding at between $\approx 7000$ and $10{,}000
\unit{km} \unit{s}^{-1}$ and possibly higher.

-- The presence of a few iron-rich plumes \citep{Fesen2015} surrounded by
shells of calcium and magnesium rich ejecta seen in the STIS spectra is in
line with delayed-detonation SNe~Ia models.  The roughly spherical distribution
of Ca and Mg argues against a highly asymmetric explosion like that 
from the merger of two white dwarfs.  {\sl HST} images and the new spectral
data, coupled with historical data on SN~1885, point to a subluminous,
delayed-detonation SN~Ia event similar to SN~1986G.

These spectra, along with previous {\sl HST} images of SNR~1885, provide
guideposts for future modeling of SN~Ia explosions, especially
subluminous Type Ia events.  

Less certain is the meaning of SN~1885's properties to other relatively young
SNe~Ia remnants.  In the case of the Galactic Type Ia supernova
of 1006 AD, optical and UV spectra of background sources show strong UV
\ion{Fe}{2} absorption but no \ion{Ca}{2} H \& K absorption
\citep{Winkler2005}. SNR~1885, in contrast, exhibits nearly saturated
\ion{Ca}{2}.  Absorption studies of other young SN~Ia remnants
could investigate elemental distributions across a wider range of SNe~Ia
events, leading to a better understanding of these important stellar explosions.

\acknowledgments
This research was based on NASA/ESA Hubble Space Telescope program GO-13471 from
the Space Telescope Science Institute, which is operated by the Association of
Universities for Research in Astronomy, Inc. under NASA contract No.
NAS5-26555.


{}


\begin{thebibliography}{}
\bibitem[Bloom et al.(2012)]{Bloom2012} Bloom, J.~S., Kasen, D., 
         Shen, K.~J., et al.\ 2012, \apjl, 744, L17 
\bibitem[Burstein et al.(1988)]{Burstein1988} Burstein, D., Bertola, 
         F., Buson, L.~M., Faber, S.~M., \& Lauer, T.~R.\ 1988, \apj, 328, 440 
\bibitem[Chevalier \& Plait(1988)]{CP88} Chevalier, R.~A., \& Plait, P.~C.\ 1988, \apjl, 331, L109
\bibitem[Colgate \& McKee(1969)]{CK69} Colgate, S.~A., \& McKee, C.\ 1969, \apj, 157, 623 
\bibitem[Dalcanton et al.(2012)]{Dalcanton2012} Dalcanton, J.~J., Williams, 
         B.~F., Lang, D., et al.\ 2012, \apjs, 200, 18 
\bibitem[de Vaucouleurs \& Corwin(1985)]{deV85} de Vaucouleurs, G., \& 
         Corwin, H.~G., Jr.\ 1985, \apj, 295, 287
\bibitem[Di Stefano \& Kilic(2012)]{Stefano2012} Di Stefano, R., \& Kilic, M.\ 2012, \apj, 759, 56
\bibitem[Diamond et al.(2015)]{Diamond15} Diamond, T.~R., Hoeflich, P., \& 
         Gerardy, C.~L.\ 2015, \apj, 806, 107 
\bibitem[Ferguson \& Davidsen(1993)]{FD93} Ferguson, H. C., \& Davidsen, A. F. 1993, \apj, 408, 92
\bibitem[Fesen et al.(1999)]{Fesen1999} Fesen, R.~A., Gerardy, C.~L.,
         McLin, K.~M., \& Hamilton, A.~J.~S.\ 1999, \apj, 514, 195
\bibitem[Fesen et al.(2015)]{Fesen2015} Fesen, R.~A., H{\"o}flich, P.~A., \& Hamilton,
                 A.~J.~S.\ 2015, \apj, 804, 140
\bibitem[Fesen et al.(2007)]{Fesen2007} Fesen, R.~A., H{\"o}flich, P.~A., Hamilton, 
         A.~J.~S., Hammell, M.~C., Gerardy, C.~L., Khokhlov, A.~M., \& 
         Wheeler, J.~C.\ 2007, \apj, 658, 396 
\bibitem[Fesen et al.(1989)]{Fesen89} Fesen, R.~A., Saken, J.~M.,
        \& Hamilton, A.~J.~S.\ 1989, \apjl, 341, L55
\bibitem[Gamezo et al.(2003)]{gamezo03} Gamezo, V.~N., Khokhlov, 
         A.~M., Oran, E.~S., Chtchelkanova, A.~Y., \& Rosenberg, R.~O.\ 2003, Science, 299, 77 
\bibitem[Gamezo et al.(2004)]{gamezo2004} Gamezo, V.~N., Khokhlov, 
         A.~M., \& Oran, E.~S.\ 2004, Physical Review Letters, 92, 211102 
\bibitem[Gamezo et al.(2005)]{gamezo2005} Gamezo, V.~N., Khokhlov, 
         A.~M., \& Oran, E.~S.\ 2005, \apj, 623, 337 
\bibitem[Garcia-Senz \& Woosley(1995)]{Garcia1995} Garcia-Senz, D., 
         \& Woosley, S.~E.\ 1995, \apj, 454, 895 
\bibitem[Hamilton \& Fesen(1988)]{HF88} Hamilton, A. J. S., \& Fesen, R. A.  1988, \apj, 327, 178
\bibitem[Hamilton \& Fesen(1991)]{HF91} Hamilton, A.~J.~S., \& Fesen, R.~A.\ 1991, in Supernovae,
         10th Santa Cruz Summer Workshop in Astronomy and Astrophysics,
         ed.\ S.~E.~Woosley (Berlin: Springer-Verlag), 656
\bibitem[Hamilton \& Fesen(2000)]{Hamilton00} Hamilton, A.~J.~S., \& Fesen, R.~A.\ 2000, 
         \apj, 542, 779
\bibitem[Hillebrandt \& Niemeyer(2000)]{HN00} Hillebrandt, W., \& 
         Niemeyer, J.~C.\ 2000, \araa, 38, 191
\bibitem[H{\"o}flich et al.(2013)] {Hoeflich2013} H{\"o}flich, P.,
         Dragulin, P., Mitchell, J., Penney, B., Sadler, B., Diamond, T.,
         Gerardy, C. \ 2013, Frontiers of Physics, 8, 144
\bibitem[H{\"o}flich et al.(2002)]{Hoeflich02} H{\"o}flich, P., Gerardy, C.~L., Fesen, R.~A., 
        \& Sakai, S.\ 2002, \apj, 568, 791 
\bibitem[H{\"o}flich et al.(2004)]{Hoeflich2004} H{\"o}flich, P.,
         Gerardy, C.~L., Nomoto, K., Motohara, K., Fesen, R.~A., Maeda, K., Ohkubo,
         T., \& Tominaga, N.\ 2004, \apj, 617, 1258
\bibitem[H{\"o}flich \& Khokhlov(1996)]{hk96} {H{\"o}flich}, P., 
         {Khokhlov}, A. 1996, \apj, 457, 500
\bibitem[H{\"o}flich \& Stein(2002)]{hs02} H{\"o}flich P., \&  Stein J.\ 2002, \apj, 568, 771
\bibitem[H{\"o}flich, Wheeler \& Thielemann(1998)]{HWT98} H{\"o}flich, P., 
         Wheeler, J. C., and Thielemann, F.K. 1998, \apj, 495, 617
\bibitem[Hofmann et al.(2013)]{Hof13} Hofmann, F., Pietsch, W.,
         Henze, M., et al.\ 2013, \aap, 555, A65
\bibitem[Howell(2011)]{Howell2011} Howell, D.~A.\ 2011, Nature Communications, 2, 350 
\bibitem[Hoyle \& Fowler(1960)]{hf60} Hoyle, F., \& Fowler, W.~A.\ 1960, \apj, 132, 565
\bibitem[Huggins(1886)]{Huggins1886} Huggins, W. 1886, \nat, 32, 465
\bibitem[Jeffery(1993)]{Jeffery1993} Jeffery, D.~J.\ 1993, \apj, 415, 734 
\bibitem[Khokhlov(1991)]{khokhlov91} Khokhlov, A.~M.\ 1991, \aap, 245, L25 
\bibitem[Khokhlov(1995)]{khokhlov95} Khokhlov, A.~M.\ 1995, \apj, 449, 695
\bibitem[Li et al.(2003)]{li03} Li, W., Filippenko, A. V., Chornock, R., 
         et al. 2003, \pasp, 115, 453L
\bibitem[Livne(1999)]{Livne99} Livne, E.\ 1999, \apjl, 527, L97
\bibitem[Livne et al.(2005)]{Livne05} Livne, E., Asida, S. M., H{\"o}flich, P.\ 2005, \apj, 632, 443
\bibitem[Maeda \& Terada(2016)]{Maeda2016} Maeda, K., \& Terada, Y.\ 2016, 
          International Journal of Modern Physics D, 25, 1630024 
\bibitem[Maunder(1885)]{Maunder1885} Maunder, E.~W.\ 1885, The Observatory, 8, 321 
\bibitem[Maunder(1886)]{Maunder1886} Maunder, E.~W.\ 1886, 
         Memorie della Societa Degli Spettroscopisti Italiani, 14, 144 
\bibitem[McConnachie et al.(2005)]{McConn2005} McConnachie, A.~W., 
         Irwin, M.~J., Ferguson, A.~M.~N., et al.\ 2005, \mnras, 356, 979 
\bibitem[Morton(1991)]{Morton91} Morton, D.~C.\ 1991, \apjs, 77, 119 
\bibitem[Niemeyer \& Hillebrandt(1995)]{Neimeyer95} Niemeyer, J.~C., \& 
         Hillebrandt, W.\ 1995, \apj, 452, 779 
\bibitem[Niemeyer et al.(1996)]{Niemeyer1996} Niemeyer, J.~C., 
         Hillebrandt, W., \& Woosley, S.~E.\ 1996, \apj, 471, 903 
\bibitem[Nomoto(1982)]{Nomoto82} Nomoto, K.\ 1982, \apj, 253, 798 
\bibitem[Nomoto et al.(1984)]{Nomoto84} Nomoto, K., Thielemann, F.-K., \& 
         Yokoi, K.\ 1984, \apj, 286, 644 
\bibitem[Nugent et al.(2011)]{Nugent2011} Nugent, P.~E., Sullivan, M., 
         Cenko, S.~B., et al.\ 2011, \nat, 480, 344 
\bibitem[Pastorello et al.(2008)]{Pastorello08} Pastorello, A., et al.\ 2008, \mnras, 389, 113
\bibitem[Penney et al.(2014)]{Penney2014} Penney, R., H{\"o}flich, P., 2014, \apj 795, 84 
\bibitem[Perets et al.(2011)]{Perets2011} Perets, H.~B., Badenes, 
         C., Arcavi, I., Simon, J.~D., \& Gal-yam, A.\ 2011, \apj, 730, 89 
\bibitem[Reinecke et al.(1999)]{Reinecke99} Reinecke, M., Hillebrandt, W., \& 
         Niemeyer, J.~C.\ 1999, \aap, 347, 739 
\bibitem[Reinecke et al.(2002)]{r02} Reinecke, M., Hillebrandt, W., \& Niemeyer, J.~C.\ 2002, \aap, 391, 1167 
\bibitem[R{\"o}pke et al.(2012)]{roepke12} {R{\"o}pke}, F.K., Kromer, D.,
         {Seitenzahl}, I.R. et al. 2012, ApJLet 750, L19
\bibitem[Sjouwerman \& Dickel(2001)]{SD01} Sjouwerman, L.~O., \& Dickel, J.~R.\ 2001, 
         Young Supernova Remnants, 565, 433 
\bibitem[Sugimoto \& Nomoto(1980)]{Sugimoto80} Sugimoto, D., \& Nomoto, K.\ 1980, \ssr, 25, 155 
\bibitem[Timmes \& Woosley(1992)]{Timmes1992} Timmes, F.~X., \& 
         Woosley, S.~E.\ 1992, \apj, 396, 649 
\bibitem[Winkler et al.(2005)]{Winkler2005} Winkler, P.~F., Long, K.~S., 
         Hamilton, A.~J.~S., \& Fesen, R.~A.\ 2005, \apj, 624, 189 
\end{thebibliography}
\end{document}